\documentclass[a4paper,fleqn,usenatbib]{mnras}
\usepackage{hyperref}

\usepackage{rotating}
\usepackage{caption}
\usepackage{lscape}

\usepackage{graphicx}	
\usepackage{amsmath}	
\usepackage{amssymb}	
\usepackage{bm}		

\usepackage{txfonts} 
\usepackage{color}

\newcommand{\kms}{km\,s$^{-1}$} 

\newcommand{\nai}{\ion{Na}{i}}
\newcommand{\hii}{\ion{H}{ii}}

\newcommand{\nii}{[\ion{N}{ii}]}
\newcommand{\oi}{[\ion{O}{i}]}
\newcommand{\oii}{[\ion{O}{ii}]}
\newcommand{\oiii}{[\ion{O}{iii}]}
\newcommand{\sii}{[\ion{S}{ii}]}

\newcommand{\ha}{H$\alpha$} 
\newcommand{\hb}{H$\beta$} 

\newcommand{\SN}{S$/$N}

\title{Systematic study of outflows in the Local Universe using CALIFA: I. Sample selection and main properties.}

\author[C.~L\'opez-Cob\'a et al.]{Carlos~L\'opez-Cob\'a$^{1}$\thanks{clopez@astro.unam.mx}, Sebasti\'an~F.~S\'anchez$^{1}$,
Joss~Bland-Hawthorn$^{2,3}$, \newauthor
Alexei~V.~Moiseev $^{4,5,6}$,
Irene~Cruz-Gonz\'alez$^{1}$, 	
Rub\'en~Garc\'ia-Benito$^{7}$,\newauthor
Jorge~K.~Barrera-Ballesteros$^{1,8}$
Llu\'is~Galbany$^{9}$
\\
$^{1}$Instituto de Astronom\'ia, Universidad Nacional Aut\'onoma de  M\'exico, Circuito Exterior, Ciudad Universitaria, Ciudad de M\'exico 04510,  Mexico \\
$^{2}$Sydney Institute for Astronomy, School of Physics, University of Sydney, NSW 2006, Australia \\
$^{3}$Centre of Excellence for All Sky Astrophysics in 3D (ASTRO-3D), Australia \\
$^{4}$ Special Astrophysical Observatory, Russian Academy of Sciences, Nizhny Arkhyz 369167, Russia \\
$^{5}$ Lomonosov Moscow State University, Sternberg Astronomical Institute, Universitetsky pr. 13, Moscow 119234, Russia\\
$^{6}$ Space Research Institute, Russian Academy of Sciences, Profsoyuznaya ul. 84/32, Moscow 117997, Russia  \\
$^{7}$ Instituto de Astrof\'sica de Andaluc\'ia (IAA/CSIC), Glorieta de la Astronom\'ia s/n Apdo. 3004, 18080, Granada, Spain\\
$^{8}$ Department of Physics \& Astronomy, John Hopkins University, Bloomberg Center, 3400 N. Charles St., Baltimore, MD 21218, USA\\
$^{9}$ PITT PACC, Department of Physics and Astronomy, University of Pittsburgh, Pittsburgh, PA 15260, USA
}

\date{Accepted XXX. Received YYY; in original form ZZZ}

\pubyear{2018}

\hypersetup{draft}
\begin{document}
\label{firstpage}
\pagerange{\pageref{firstpage}--\pageref{lastpage}}
\maketitle

\begin{abstract}
  
We present a sample of 17 objects from the CALIFA survey where we find
initial evidence of galactic winds based on their off-axis ionization
properties.  We identify the presence of outflows using various
optical diagnostic diagrams (e.g., EW(H$\alpha$), \nii/\ha, \sii/\ha,
\oi/\ha\ line-ratio maps).  We find that all 17 candidate outflow
galaxies lie along the sequence of active star formation in the
M$_\star$ vs. star-formation rate diagram, without a clear excess in
the integrated SFR. The location of galaxies along the star-formation
main sequence (SFMS) does not influence strongly the presence or not
of outflows. The analysis of the star-formation rate density
($\Sigma_{\rm SFR}$) reveals that the CALIFA sources present higher
values when compared with normal star-forming galaxies. The strength
of this relation depends on the calibrator used to estimate the
SFR. This excess in $\Sigma_{\rm SFR}$ is significant within the first
effective radius supporting the idea that most outflows are driven by
processes in the inner regions of a galaxy.  We find that the
molecular gas mass density ($\Sigma_\mathrm{gas}$) is a key parameter
that plays an important role in the generation of outflows through its
association with the local SFR. The canonical threshold reported for
the generation of outflows -- $\Sigma_{\rm SFR}>0.1$ $\mathrm{M}_\odot
\mathrm{yr}^{-1} \mathrm{kpc}^{-2}$ -- is only marginally exceeded in
our sample.  Within the Kennicutt-Schmidt diagram we propose a domain
for galaxies hosting starburst-driven outflows defined by $\Sigma_{\rm
  SFR}>10^{-2} \,\mathrm{M}_\odot \mathrm{yr}^{-1} \mathrm{kpc}^{-2}$
and $\Sigma_\mathrm{gas}>10^{1.2} \, \mathrm{M}_\odot
\mathrm{pc}^{-2}$ within a central kiloparcec region.

\end{abstract}


\section{Introduction}
Galactic outflows have been invoked in many astrophysical problems to explain some local and global properties of galaxies like the tight correlation in the stellar mass and metallicity \citep[][]{Tremonti2004,Heckmanetal2002}, the metal enrichment of intergalactic medium \citep[][]{Pettini1998,Veilleuxetal2005}, and in the current models of galaxy formation, where the amount of feedback from outflows is a key ingredient  which is not well constrained \citep[e.g,][]{Aguirre2001,Springel2003,Scannapieco2006}. Even their global effect, either in preventing or triggering star-formation is under discussion \citep[e.g.,][]{silk13}.

Outflows are driven either by supernovae explosions (SN), stellar winds or by active galactic nuclei (AGN), or some combination of these $-$ we refer to such objects collectively as active galaxies. Nuclear star formation is found to occur in the nuclear regions of most Seyfert galaxies \citep{Esquej2014} and so both may act in concert to generate outflows, although just how this works is mysterious \citep{Hopkins2010}.
The scale of these outflows depends partly on the escape velocity via the gravitational potential well \citep{Tanner2017}. A high fraction of active galaxies with lower total mass are expected to host outflows because of their lower escape velocity \citep{Martin1998,BlandHawthorn2015}. This favours the loss of large fractions of gas and metals in these galaxies \citep[e.g.,][]{Barrera-Ballesteros2018}. Massive galaxies retain their baryons more effectively although considerable recycling throughout the halo can take place \citep[]{Cooper2008,Tanner2016}.

Although there has been extensive effort on the theory of outflows, the observational counterparts are far from being understood, mainly because their multiphase nature makes them hard to detect and interpret. Numerical simulations today are far from capturing the full complexity of galactic winds too \citep[e.g.,][]{mart11}.
Outflows have been detected at high redshift \citep{Coil2011,Genzel2014}, in the nearby universe \citep[e.g.,][]{Franx1997,Fogarty2012,Ho2014,Ho2016}, even in the Local Group \citep{Blandhawthorn2003,Su2010,Fox2015}.
Galactic winds have also been detected across most galaxy types \citep[e.g.,][]{Axon1978,Bland1988,Heckman1990,Lehnert1996,Martin1998,Rupke2005a}. Despite of all these studies, the nature, properties and influence of outflows in galaxy evolution are still unclear.
How, why and where outflows are produced in a galaxy, as well as the loss rates of mass, metals and energy that they produce are still open questions that have not been completely  understood \citep[see][for an extensive review]{Veilleuxetal2005}. High quality spatially resolved spectroscopic data could bring some light in the understanding of these processes. 

A primary problem when studying outflows is the detection itself. Outflows are commonly studied in starburst galaxies or in ultraluminous infrared galaxies (ULIRG) due to their large star formation rates (SFR), making them more prone to develop outflows. This means that studies of outflows are biased towards galaxies with high star formation rates. Other studies analyze outflows directly associated with strong AGNs, in particular with those ones directly pointing towards the observer \citep[e.g., BL Lacs or Blazars,][]{Antonucci1985,Scarpa2000,Celotti2008}, being biased towards these particular kind of objects. Thus, there are few systematic studies of the presence of outflows in an unbiased population of galaxies \citep[cf.,][]{Sharp2010,Ho2016}.

Early studies of galaxies with {\it bona fide} outflows have constructed the basis in the methodology to detect and characterize them \citep[e.g.,][]{Heckman1990,Lehnert1996,Rupke2005a,Rupke2005b}.
This methodology is based on the study of the spatial distribution of certain emission line ratios over the extra-planar regions of disk galaxies, and their comparison with certain kinematic properties. Although these studies provide moderate samples of outflows, they do not provide well defined statistics about the frequency of galaxies hosting outflows and their properties in comparison with those not hosting them. In the present study, we address the search and characterization of the statistical properties of galaxies hosting outflows. For this purpose, we exploit the CALIFA integral field spectroscopic survey (IFS) which achieved a large sample (835) of galaxies observed from the Calar Alto telescope in Spain.

Different optical IFS surveys \citep[e.g., SAMI, MaNGA, AMUSING,][]{SAMI,ManGA,Galbany2016},
have already taken advantage of this technique to study the spatially resolved properties of the warm ionized gas component of outflows \citep[e.g.,][]{Sharp2010,Rich2011,Fogarty2012, Ho2014,Wild2014,Rich2015,Ho2016,prieto16,Coba2017,Maiolino2017}. The use of larger samples allows us to perform statistical analysis not only on the outflowing galaxies, but on those that do not present outflows, i.e., a properly selected control sample typically overlooked in resolved galaxy surveys.

The layout of this article is as follows: In Sec. \ref{data}, we present the data and physical properties (data products) extracted from them used along this article; describing the analysis of the stellar population in Section \ref{ana_ssp} and of the ionized gas in \ref{ana_ion}. The  outflow sample analysis of these data products is presented in Section \ref{ana}.  It includes the selection of candidates that host outflows in Sec. \ref{select}, and a description of their distribution along the color-magnitude diagram (CMD), in Sec. \ref{sec:CMD}, their masses and morphologies, in Sec. \ref{sec:morph} and source of the ionization in the central regions, in Sec. \ref{sec:cen_ion}.  All of these properties are presented in comparison with those of galaxies without a host outflow. The comparison of the integrated SFR (Sec. \ref{sec:SFMS}), the radial distribution of the SFR density (Sec. \ref{sec:rad}), and their central values (Sec. \ref{sec:sig_SFR}), have lead to the main results of this investigation, discussed in Section \ref{sec:res}. The conclusions and future perspectives are presented in Sec. \ref{sec:con}.  In this work the standard $\Lambda$CDM cosmology with H$_0$ = 70 \kms\,Mpc$^{-1}$, $\Omega_\mathrm{m}=0.3$, $\Omega_\mathrm{\Lambda}=0.7$ is adopted.

\section{Data cubes and data products}
\label{data} 

The analysed sample comprises all galaxies in the CALIFA survey\footnote{\url{http://califa.caha.es}} \citep[e.g.,][]{CALIFA1} up to January 2018, i.e., those with good quality spectroscopic data observed at the 3.5-m of Centro Astron\'omico Hispano-Alem\'an (CAHA). It includes the 667 galaxies comprising the 3rd CALIFA Data Release \citep[e.g.,][]{CALIFA3}, and in addition we include those galaxies with good quality data that were excluded from DR3 because either they did not have SDSS-DR7 imaging data (a primary selection for DR3) or they were observed after the final sample was closed, as part of the CALIFA-extended programs \citetext{e.g., \citealp{Garcia-Benito2017}; PISCO: \citealp{PISCO}}. The final sample comprises a total of 835 galaxies. All galaxies were selected following the same primary selection criteria of the main CALIFA survey, i.e., that their optical extent fits within the field-of-view (FoV) of the instrument, relaxing other selection criteria outlined in \citet{walcher14} like the redshift range or the absolute magnitude. Thus, this compilation is essentially a diameter-selected survey.

The details of the CALIFA survey, including the observational strategy and data reduction are explained in \citet{CALIFA1} and \citet{CALIFA3}. All galaxies were
observed using PMAS \citep[e.g.,][]{roth05} in the PPaK configuration
\citep[e.g.,][]{kelz06}, covering an hexagonal field of view (FoV) of
74$\arcsec$$\times$64$\arcsec$, which is sufficient to map the full optical extent of the galaxies up to two to three disk effective radii. This is possible because of the diameter selection of the CALIFA sample \citep[e.g.,][]{walcher14}. The observing strategy guarantees complete 
coverage of the FoV, with a final spatial resolution of
FWHM$\sim$2.5$\arcsec$, corresponding to $\sim$1 kpc at the average redshift of the survey \citep[e.g,][]{CALIFA2,CALIFA3}. The sampled wavelength range and
spectroscopic resolution for the adopted setup (3745-7500 \AA,
$\lambda/\Delta\lambda\sim$850, V500 setup) are more than sufficient
to explore the most prominent ionized gas emission lines from
\oii$\lambda$3727 to \sii$\lambda$6731 at the
redshift of our targets,
and to deblend and subtract the underlying stellar population 
\citep[e.g.,][]{kehrig12,cid-fernandes13,cid-fernandes14,sanchez13,sanchez14,Pipe3D_I}.
The current dataset was
reduced using version 2.2 of the CALIFA pipeline, whose modifications with respect to previous ones \citep[e.g.,][]{CALIFA1,husemann13,CALIFA2} are described in \citet{CALIFA3}. The final product of the reduction is a datacube comprising the spatial information in the {\it x} and {\it y} axis, and the spectral one in the {\it z} axis. For further details of the adopted data-format and the quality of the data see \citet{CALIFA3}.  

\subsection{Stellar population analysis}
\label{ana_ssp}

The datacubes were analysed using the {\sc Pipe3D} pipeline \citep[e.g.,][]{Pipe3D_I}. {\sc Pipe3D} performs a combination of multiple synthetic stellar population (SSP) templates, extracted from the MILES \citep[e.g.,][]{Sanchez-Blazquez2006,Vazdekis2010,Falcon2011} and the gsd156 library \citep[e.g.,][]{Cid2013}, to determine the best stellar model. These templates cover a wide range in metallicities
from sub solar to supra solar, with different stellar ages from 1 Myr to 14 Gyr. 
Before  starting with the fitting process, a tesselation procedure is performed on the data cube in order to increase the signal-to-noise (\SN) of the stellar continuum. This segmentation produces tesselas of different sizes to achieve the desired \SN. All the spectra in each spatial bin is co-added and is treated as individual spectra, and at the end of the fitting analysis, a dezonification of the coadded spectra is applied by taking into account the area of each tessela \citep[see ][]{Cid2013}.  
A two dimensional set of data-products, described in \citet{Pipe3D_II}, are obtained from the SSP fitting.
One of such data products
is the cumulative  stellar mass at different epochs.
The stellar mass (M$_{\star}$) of a galaxy is estimated by adding the mass in each bin from the tesselation procedure,  taking into account the local luminosity of each spectrum and the mass-to-light ratio \citep[see][]{Rosa2015}.
For a given age (that defines a look-back time), the stellar mass is
\begin{equation}
M_{\star, age}=\sum_{j=1}^n M_j
\end{equation}
where the $j$ index runs over the number of templates in the SSP library up to the considered {\it age}. Integrated over all the complete set of SSP-templates it provides the actual stellar mass of the galaxy. As shown in \citet{Pipe3D_II} and Bitsakis et al. (in prep.), this stellar mass is totally consistent with the one provided using multi-band photometric data.

Having estimated the stellar mass at a certain look-back time, it is straightforward to estimate the star-formation rate at this particular time. The SFR would be the differential mass at two adjacent times ($\Delta t_{age}$) over the time range between them $\Delta t_{age}$:
\begin{equation}
SFR_{age}=\frac{\Delta M_{age}}{\Delta t_{age}}
\end{equation}
In \citet{Rosa2017} and Sanchez et al. (in prep.) it has been explicitly shown that this star-formation rate, that we will define as SFR$_{\rm SSP}$, correlates very well with other estimations of the SFR.

\begin{figure*}[h]
\centering
\includegraphics[width=\textwidth]{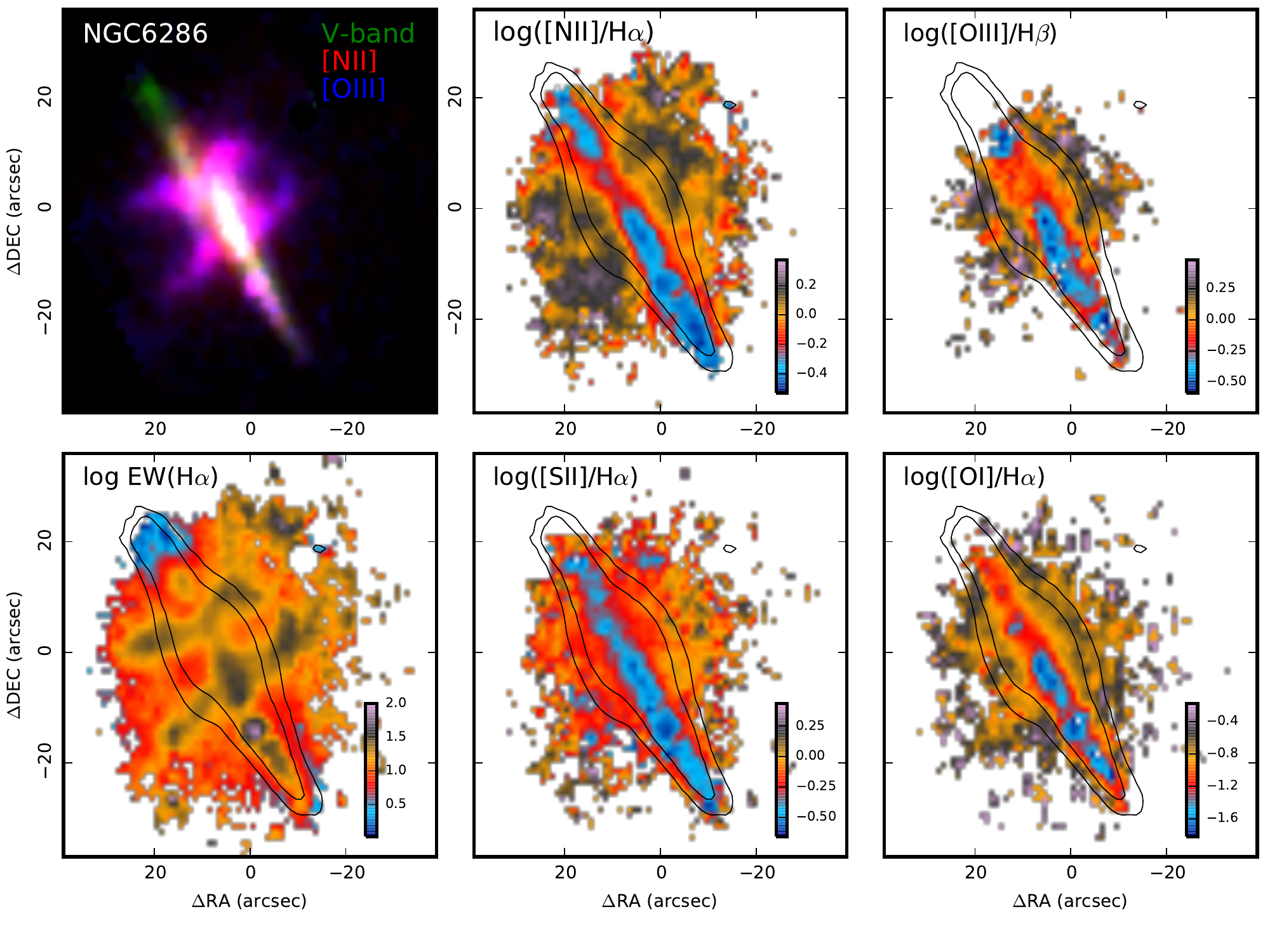}
\caption{Example of a wind galaxy selected from the CALIFA sample with the high inclination and line-ratio criteria. This galaxy, NGC\,6286, is part of the candidates galaxies with a host outflow.  The {\it top left panel,} shows the RGB image of NGC\,6286, where red is \nii, green is the V-band and blue is \oiii. The {\it top central panel,} is the spatially resolved \nii/\ha\ line ratio map. The black contour in this, and in the others maps, indicate the continuum level at 0.1 and 0.05 $\times 10^{-16}$ erg s$^{-1}$ cm$^{-2}$, while the intensity color bar is in the right corner of each map. The {\it top right panel,} is the spatially resolved \oiii/\hb\ line ratio map. The
 {bottom left panel,} shows the 2D-Equivalent Width of \ha\ estimated with the SSP fitting analysis. The {\it bottom central panel,} shows the \sii$\lambda\lambda6717,6731$/\ha\ line ratio map; and the {\it bottom right panel,} the \oi$\lambda6300$/\ha\ line ratio map.}
\label{example_ic2101}
\end{figure*}

\begin{figure*}[h]
\centering
\includegraphics[width=\textwidth]{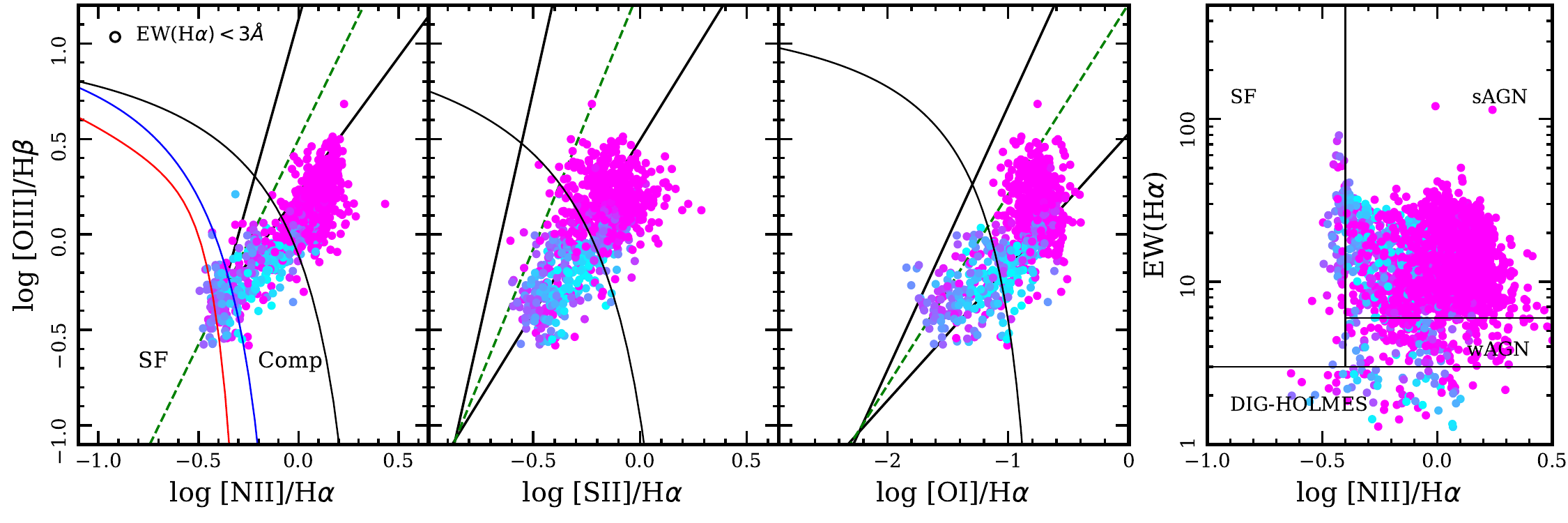}
\caption{Diagnostic diagrams for the galaxy NGC\,6286. All the emission lines involved have a \SN\ $>2$. {\it Cyan} dots represent spaxels lying in the disk while {\it pink} dots represent spaxels lying beyond { 5\arcsec} from the disk, i.e., in the extraplanar region.  The dark blue colors in this colormap represent the transition from disk to the extraplanar region. This color scheme is adopted in all the candidates regardless of their inclination. The black continuous curve in the first three panels represent the \citet{Kewley2001} demarcations from  SF regions and AGNs. The red and blue lines in the first panel represent the \citet{Kauffmann2003} and \citet{Stasinka2006} curves respectively. The black circle represent those spaxels in which EW(\ha) $<$ 3 \AA. For this particular case, all the spaxels have EW(\ha) $>$ 3 \AA. The continuous diagonal black lines represent the locus of shock ionization (rightmost) and AGN ionization (leftmost) 
calibrated with bonafide outflows dominated by shock-excited emission and AGN-excited emission
according to \citet{Sharp2010}. The green dashed line between represent the bisector line between the shocks and AGN ionization. 
The right panel shows the WHAN diagram \citep{Cid2011}, \nii/\ha\ vs. EW(\ha). } 
\label{DD_ic0480}
\end{figure*}

\subsection{Ionized gas analysis}
\label{ana_ion}

Once obtained the best stellar population model for each spectra in the data cube, it is subtracted from the original cube to obtain a pure-gas cube, following the procedures described in \citet{Pipe3D_I}. Then, we analysed each of the detected emission lines in each individual spectrum within this cube using the fitting code {\sc fit3D} \citep[e.g.,][]{Pipe3D_II}. For this particular study it was performed a  non parametric method based on a  moment analysis in the  pure-gas cube  as described in \citet{Pipe3D_II}. We recover the main properties of the emission lines, including the integrated flux intensity, line velocity and velocity dispersion. For this analysis, we assume that all emission lines within a spaxel share
the same velocity and velocity dispersion. The result of this procedure applied to each data cube is a set of bi-dimensional maps of the considered parameters, with their corresponding errors,  for each analysed emission line.

In addition to these parameters the equivalent width (EW) of each emission line  is derived. In particular that of \ha\, which will be used in our scheme of classification of the ionization source. 
To derive this quantity  the stellar continuum flux density is estimated prior to the subtraction of the stellar model. Then, the integrated flux of each emission line, derived by the  moment analysis is divided by this continuum density, at the wavelength of the emission lines, resulting in the required EW. 

\section{Outflow sample analysis}
\label{ana}

\subsection{Candidates selection}
\label{select}

 Highly inclined galaxies are particularly good candidates to detect extraplanar ionized gas and therefore they are more suitable candidates to host outflows. We started the selection process by considering only those galaxies with high inclination ($i>$ 70$^{\circ}$),  in order to minimize the effect of mixing of ionization along the line-of-sight due to projection effects. Although we cannot preclude for a certain level of contamination. 
Using this criterion results in 203 galaxies. Then, we select those galaxies with an
increase in the optical line ratios \nii/\ha, \oiii/\hb, \sii/\ha\ and \oi/\ha\ along the semi-minor axis and the disk vertical direction.  These increments are characteristic of shocks produced by galactic outflows \citep{Veilleux2002,Veilleuxetal2005}, although  they are not exclusive of these processes.  Here increase means that the ionization is not compatible with the typical line-ratios observed in SF regions. Outflows are favoured to expand in the direction of the lowest gradient of pressure, which is found along the semi-minor axis or  in the extraplanar region. As larger the inclination, sharper will be observed the separation between the soft ionization from the SF regions, and the harder  would be the ionization produced by shocks in outflows. 

Within the high inclined sub-sample we find that 39 galaxies present such line ratios enhancement. In Figure \ref{example_ic2101} we present one galaxy, NGC\,6286, that  complies with  these criteria. In this figure it is clear how extraplanar gas extends beyond the continuum extension. Hereafter, we will refer to the {\it disk} region, regardless of the inclination, as the area located within $\pm$\,5\,arcsec\, from the semi-major axis, and as {\it extraplanar} region to the area located  farther than 5\,\arcsec\, of this axis.  This transition region varies in each galaxy, although it represents a mean value at the average redshift of CALIFA ($\sim 2$ kpc). NGC\,6286 clearly shows ionized gas in the extraplanar region, with larger line ratios than the ones detected in the disk. Even more, this galaxy presents the  archetypal biconical distribution expected  in an outflow in both the emission line intensities and  ratios. However, this  morphological structure does not exist in all outflows. In many cases they present a variety of complex morphologies \citep[e.g.,][]{Cecil2001,Veilleux2001,Martin2002,Veilleux2002}. 

Depending on the shock velocities, ionization by shocks can cover a wide range of line ratios making them  rather difficult to identify in the classical diagnostic diagrams \citep[e.g.,][]{Baldwin1981,Veilleux1987}, at difference with other ionizing sources which are confined to specific regions of these diagrams. 
Therefore, it is not possible to define demarcation curves, like the ones used to separate, for example, SF and AGN-like ionization zones \citetext{K01: \citealp{Kewley2001}; K03: \citealp{Kauffmann2003}; S06: \citealp{Stasinka2006}}. Even more, shock-like ionization overlaps in these diagrams with locations of SF regions of intermediate/high metallicities, low-luminosity AGNs and/or post-AGBs ionizations \citep[e.g.,][]{Alatalo2016}. 
Despite of its complexity, there have been efforts trying to constrain the location of shocks in these diagrams, either with direct observations of bonafide outflows \citep[e.g.,][]{Sharp2010}, or by the implementation of shock models such as \textsc{mappings-iii} \citep[e.g.,][]{Dopita1995,Dopita1996,mappings}.

In addition to this problem, some characteristics of outflows, like the enhanced line ratios in the extraplanar regions and their location in the diagnostic diagrams, are shared with the ionization produced by the so-called hot low-mass evolved stars \citep[e.g., HOLMES,][]{Flores2011}. These old stars can dominate the production of ionizing photons with respect to young massive stars and produce the ionization observed in the extraplanar diffuse ionized gas (eDIG/DIG). Their effect could be particularly important in early-type galaxies \citep[e.g,][]{Binette1994,Stasinska2008}, and can reproduce the observed ionization in the so-called low-ionization nuclear emission-line regions \citep[LINER,][]{Heckman1980}, classically associated with  low-luminosity AGNs. This ionization has been recently found to be ubiquitous in the retired regions of any galaxy \citep[e.g.,][]{sign13,belf16b}, and frequently detected in edge-on galaxies \citep[e.g.,][]{Amy2017,Lacerda2018}. 
In star-forming galaxies, the contamination by DIG, regardless of the galaxy inclination, is to move the SF regions towards the composite or the LINER region in the diagnostic diagrams \citep[e.g.,][]{Zhang2017}. Therefore, if we had adopted a lower value in the inclination angle, or if we had searched for outflows regardless of their inclination, it would increase the DIG fraction in our sample. 

A characteristic that share both DIGs and HOLMES is their low equivalent width of \ha\ \citep[e.g.,][]{Stasinska2008}. As a method to distinguish this ionization from shocks, we adopted the  WHAN diagram introduced by \citet{Cid2011} which uses the \nii/\ha\ vs. equivalent width of \ha\  to distinguish between true and fake AGNs (retired galaxies with EW(\ha) $<$ 3 \AA). We impose this additional criterion to select galaxies dominated by shocks by excluding those ones in which the extraplanar ionized gas is largely dominated by regions with EW(\ha) $<$ 3 \AA\ (i.e., if they are compatible with eDIG, either post-AGBs or HOLMES ionization). 
  
 Figure~\ref{DD_ic0480} shows an example of the implementation of the classical diagnostic diagrams (\nii/\ha, \sii/\ha\ and \oi/\ha\ vs. \oiii/\hb) along with the WHAN diagram, for the
spatially resolved components,  both disk and extraplanar regions, applied to the archetypal outflow galaxy NGC\,6286. 
Although a fraction of the extraplanar gas falls  below the SF  demarcation line by K01, probably due to projection effects, and a fraction of the disk  gas falls in the sAGN region of the WHAN diagram, it is clear that the extraplanar gas is not compatible with being ionized by old stars but by a strong source of ionization. We have included in these diagrams the locus of AGN and shock-excited emission from \citet{Sharp2010} to distinguish between these two types of outflows. The location of the outflowing gas in the diagnostic diagrams is mainly distributed in the shock-excited region (to the right from the bisector line). We refine the classification proposed by \citet{Sharp2010}, imposing the condition that the ionized gas should have an EW(\ha)$>$3\AA\ in the extraplanar region to be classified as an outflow, and below this limit to be classified as DIG, as indicated before. Therefore, those spaxels above the \citet{Kewley2001} curve, at the left-size of the \citet{Sharp2010} demarcation line, in the extraplanar region, and with above the indicated EW(\ha) will be classified as AGN-driven outflows (or compatible with being ionized by an AGN). On the other hand, all those spaxels following the same criteria, but at the right-side of the \citet{Sharp2010} demarcation line, would be classified as shock ionized (or SF-driven outflows in general). In particular, for NGC\, 6286, the extraplanar gas is consistent to be ionized in 100\% of the spaxels by shocks, 0\% by AGN and 0\% by DIG according to the indicated criteria. The spatial distribution of the line ratios shown in Fig.~\ref{example_ic2101}, together with their location within the four diagrams  of Fig.~\ref{DD_ic0480}, allow us to conclude that  the observed outflow in NGC\,6286 is most probably 
a SF-driven and shock-excited  wind. Note that  the wind outflow in this galaxy was already suspected in the paper by \citet{Shalyapina2004} based on scanning Fabry--Perot interferometer observations in  \nii\ and \ha\ emission lines. Now CALIFA diagnostic diagrams give a detailed picture of the gas ionization in the butterfly--like extended nebulae of this object.

In summary, to select galaxies that host an outflow we adopt the following  selection criteria: (i) high inclined galaxies, (ii) detection of extraplanar ionized gas, (iii) identification of an enhance in the line ratios along the semi-minor axis, (iv) EW(\ha)$>$ 3 \AA\ in these regions, and finally (v) a biconical, bipolar or a symmetric morphology in the extraplanar gas, not homogeneously distributed at any galactocentric distance above the disk. By applying these criteria we ended up with 17 galaxies candidates to host an outflow.

We will refer hereafter to this sub-sample as galaxies {\it candidates} with a host outflow. We cannot firmly conclude that they host an outflow since we lack of the required high spectral resolution data to perform a detailed kinematics analysis. Thus,
we cannot resolve the asymmetries in the emission line profiles, frequently detected in outflows due to the expansion of the gas, or analyse the known correlation between the velocity dispersion and the line ratios, a unique signature of shock ionization \citep[e.g.,][]{Dopita1995,Lehnert1996,Monreal2010,Coba2017}. For instance, in the special case of NGC\,6286 the line-of-sight ionized gas velocity dispersion significantly increases outside the stellar disk \citep[according to Fig.~5 in][]{Shalyapina2004}.

In Table~\ref{table_candidates} we summarize the main properties of this sample of galaxies and in the Appendix~\ref{appendix:outlows} we present the same plots shown for NGC\,6286 (i.e., equivalent to Figs.~\ref{example_ic2101} and \ref{DD_ic0480}), for all  the outflow candidates in the CALIFA sample. In addition, we list in Table~\ref{Tab:Not_outflow} the remaining 26 galaxies that were not classified as outflows  by the imposed criteria, although they present extraplanar  diffuse ionized gas (eDIG), and in may cases, show an enhancement in the analysed line ratios. Following the criteria indicated before we list in this table the fraction of spaxels in the extraplanar region being compatible with either DIG, AGN-driven or SF-driven outflows, based on the combination of the classical diagnostic diagrams, the value of the EW(\ha), and the \citet{Kewley2001} and \citet{Sharp2010} demarcation lines. 


\begin{figure}
\includegraphics[width=\columnwidth]{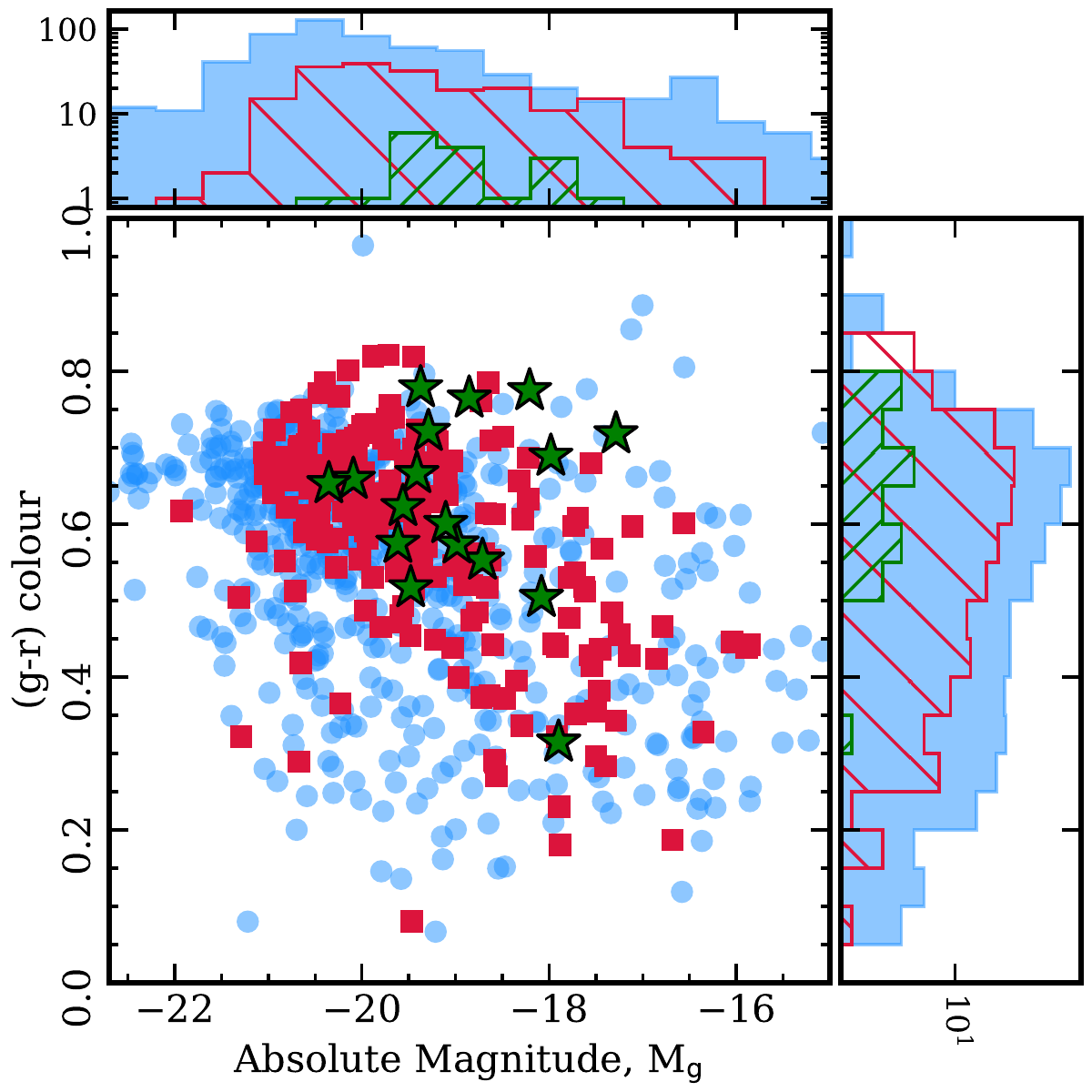}
\includegraphics[width=\columnwidth]{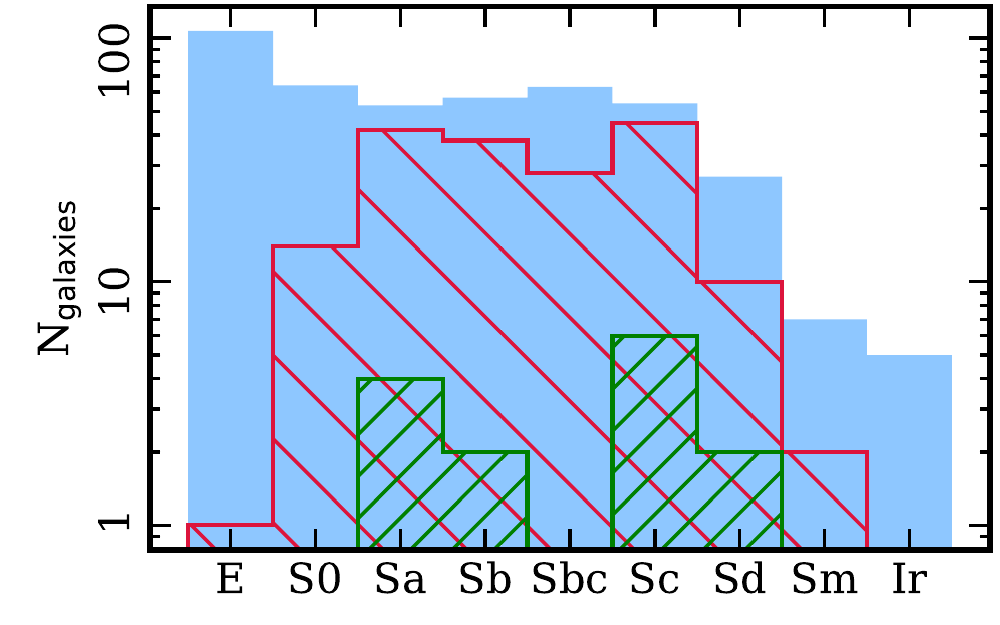}
\includegraphics[width=\columnwidth]{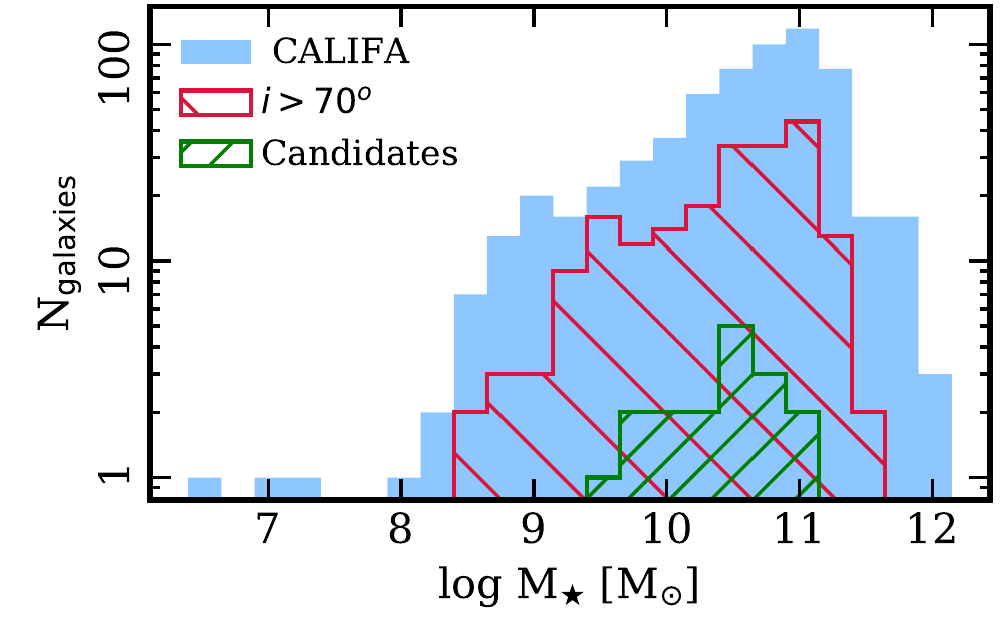}
\caption{{\it Top panel,} distribution of the three sub-samples in the $(g-r)$ vs. M$_g$ diagram. {\bf {\it Blue dots}}  represent the CALIFA sub-sample, {\bf {\it red squares}} represent the high inclination galaxies and {\bf {\it green stars}} the  outflow candidates. Histograms of the 
$(g-r)$ and M$_g$  distributions for each sub-sample are included in each axis.  The colour code of the  histograms are the same as mentioned before. {\it Middle panel:}  histogram of the morphological distribution for three sub-samples. {\it Bottom panel:} histogram of the stellar mass distribution. The morphology histogram includes the 734 galaxies observed by CALIFA up to 2017 \citep[e.g.,][]{sanchez17b} while the others two histograms include the total sample of 835 galaxies  \citep[i.e., including the PISCO sample,][]{PISCO}.} 
\label{CMD}
\end{figure}
So far, we finish our classification process with three different sub-samples: (i) Those galaxies with $i<70^\circ$, which we will denote as the CALIFA low-inclination sample, or just CALIFA sample for simplicity, since it dominates the number statistics (615 galaxies), and therefore comprises a representative sub-sample of the original one, (ii) the high-inclination galaxies (203 galaxies), and (iii) the outflow candidates (17 galaxies).

\subsection{Properties of outflow candidates}
In this section we explore the global properties of the candidates with a host outflows in comparison with those of the other two sub-samples of galaxies.
\subsubsection{The colour-magnitude diagram}
\label{sec:CMD}
Figure~\ref{CMD} shows the distribution of the three galaxy samples in the colour magnitude diagram (CMD) for the $(g-r)$ colour versus the $g$--band absolute magnitude. The  galaxies from CALIFA span over the full CMD from the red sequence to the blue cloud and over the intermediate region known as green valley, populated by transition galaxies and AGN hosts \citep[e.g.,][]{SFSanchez2017}. The global properties of the CALIFA sample has been reported in previous papers for the different data releases \citep[e.g.,][]{CALIFA1,walcher14,CALIFA2,CALIFA3}.

\begin{figure*}
\centering
\includegraphics[width=\textwidth]{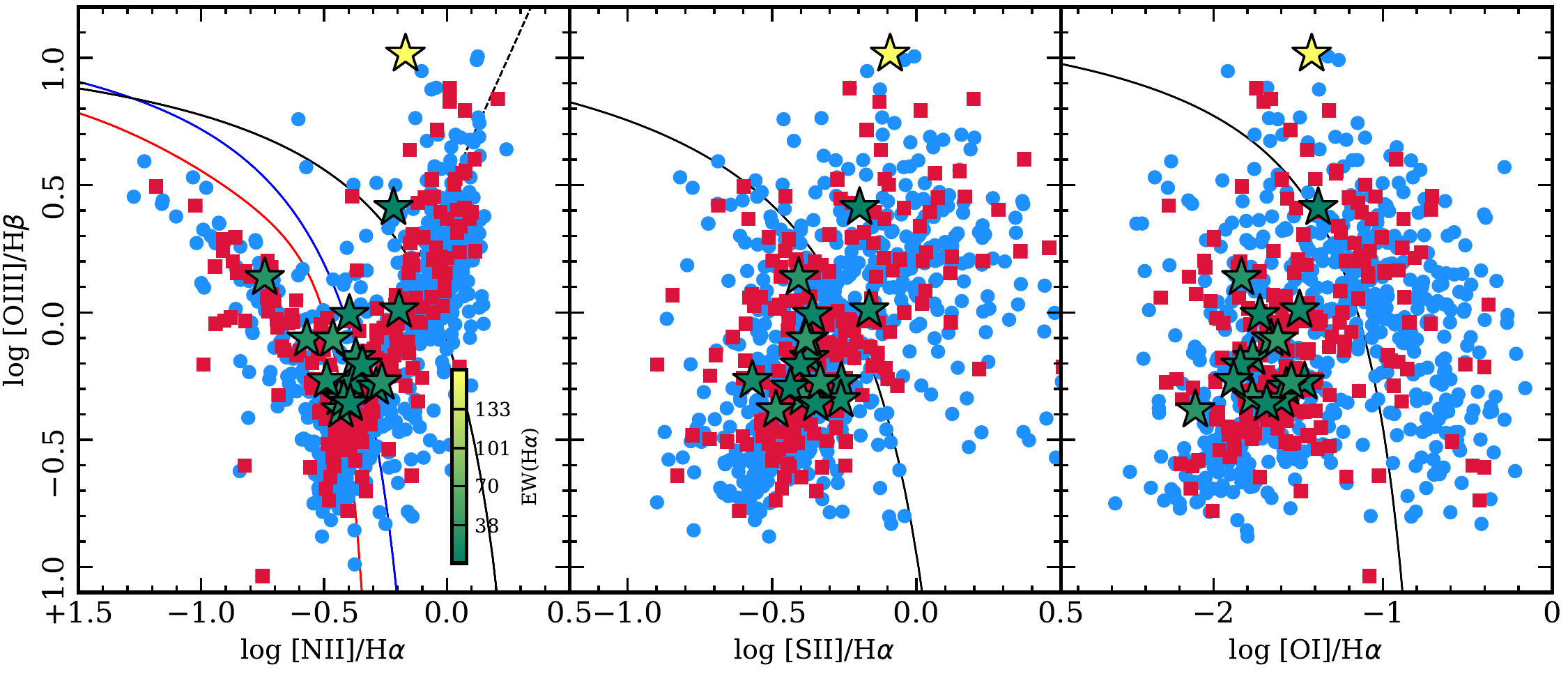}
\caption{Diagnostic diagrams for the central regions $(3\arcsec \times 3\arcsec)$ of the galaxies in the three sub-samples: the CALIFA sub-sample (cyan dots), the high inclined galaxies (red dots) and the candidates galaxies (green stars). The green colour code in the lower-right of the first panel represents the EW(\ha). We have coded only the EW(\ha) for the candidates.
The continuous black curve in the three panels represent the \citet{Kewley2001} demarcation curves. The blue and red lines in the first panel represent the demarcations from  \citet{Kauffmann2003} and \citet{Stasinka2006} respectively.
The broken line represents the demarcation between Seyfert (up-left) and LINER (up-right) from \citet{Kewley2001}. }
\label{BPT_sample}
\end{figure*}

Now, we would like to investigate the differences between the three sub-samples.
In order to quantify these differences, we performed a two dimensional Kolmogorov-Smirnov test \citep[2D KS,][]{Peacock1983,Fasano1987,Numerical_Recipes}. This test compares two 2D distributions. The null hypothesis is that the observed population of galaxies (the high inclined or the candidates) is drawn randomly from a parent population (CALIFA or the highly inclined galaxies). Typically one assumes a critical p--value to reject the null hypothesis. In our case we will adopt a p--value of $0.05$. As mentioned in \citet{Numerical_Recipes}, the resulting p--value in the 2D KS is only an approximation, and the test is accurate enough when N $\sim20$ and p--value $\lesssim 0.20$. We applied the 2D KS test for the galaxies from the CALIFA sub-sample and the highly inclined galaxies. The resulting p--value is of the order of $10^{-6}$, which is highly significant. This  implies that both samples are statistically different. Thus, the  highly inclined galaxies are not a representative sub-sample of the CALIFA galaxies, at least in the space of parameters considered. This is not really surprising, because the full CALIFA sample comprises a wide range of morphological types, with a substantial fraction of elliptical galaxies, that by definition are more roundish and prompt to be rejected from a selection of high-inclined galaxies based on the semi-major to semi-minor ratio.

We now applied the 2D KS test for the high inclination and candidate galaxies. The resulting p--value is 0.007, which is also significant at the 0.05 confidence level. In this case it is not obvious why these two samples should be so different. An inspection of Fig.~\ref{CMD} shows that the candidates occupy a narrower region in the CMD diagram, $-20.5<\mathrm{M}_g<-17.0$,  
which would probably reflect a bias in the luminosity distribution. In other words, we do not find outflow candidates brighter than $-20.5$ mag, although a 15\% of the highly-inclined galaxies are located in this range.  We also found that 5 of the outflow candidates show redder colours. These galaxies also present larger extinction values, ${\mathrm A_v}\sim 3$ mag, result of their high inclination.

\subsubsection{Morphological and stellar mass distributions}
\label{sec:morph}

Figure~\ref{CMD}, middle panel, shows the distribution in morphology and stellar mass for the three sub-samples. The CALIFA sub-sample is distributed in a wide variety of morphological types, from early- to late-types and irregulars, and stellar masses $6<\log \mathrm{\,M}_{\star}<12$. The high inclination sub-sample is clearly dominated by spiral galaxies (1 elliptical). Their mass distribution seems to follow the same as the CALIFA sub-sample, but without high/low mass galaxies, i.e., restricted to $8.5<\log \mathrm{\,M}_{\star}<11.5$. Finally, the outflow candidate sample only includes spiral galaxies of types Sa, Sb, Sc and Sd. The masses in this sub-sample are distributed in an even narrower range, $9.5<\log \mathrm{\,M}_{\star}<11$.

In order to quantify the observed differences between the mass distributions of the candidates and the  other two sub-samples, we applied the one dimensional Kolmogorov-Smirnov (K-S) test \citep[e.g.,][]{Numerical_Recipes}. The K-S test compares the maximum difference  between two cumulative distribution functions. As larger is the difference between two distributions, larger is the probability that the two distributions arise from different samples. 

With a resulting p--value of 0.02, the K-S reveals a significant difference, at the level of 0.05, between the high- and low-inclination galaxies. 
This is in concordance with the result of the 2D KS estimated for the CMD in the previous section. On the other hand, the resulting p-value for candidates and the high inclined galaxies is 0.25.  So, the mass distribution of the candidate sub-sample is consistent of being drawn from the same mass distribution of the high inclination galaxies.
Thus, the candidates present a similar stellar mass as the highly-inclined galaxies. Therefore, the differences found in the CMD are most probably due to a difference in colour, rather than in absolute magnitude (or mass). In summary, the galaxies hosting outflows seem to be slightly brighter, with a similar stellar mass and slightly more evolved stellar populations or with larger dust attenuations than the average inclined galaxies.

\subsubsection{Central ionization}
\label{sec:cen_ion}
In order to investigate the dominant ionization in the nuclear region of the galaxies from the three sub-samples, we co-add the emission line fluxes of \ha, \hb, \oiii, \nii, \sii\ and \oi\ over an area of $3''\times 3 ''$ at  the nucleus of each galaxy. Then, we plot the line ratios in the diagnostic diagrams explained before. This is shown in Fig.~\ref{BPT_sample}. 
We note that the  CALIFA and the highly inclined galaxies are distributed following the classical seagull shape, which reflects the variety of ionizing sources in these samples. 
A large fraction  of AGN populate these two sub-samples. 
The number of AGN in each sub-sample varies depending on which diagnostic is used to classify them. An AGN is classified if it lies above the K01 curve and presents an EW(\ha) $>3$ \AA\  \citep{sanchez17b}. From this we obtain that there are 61 and 16 AGN  in the CALIFA and high-inclined sub-samples respectively in the \nii/\ha\ diagram, 71 and 28 AGN in the \sii/\ha\ diagram, and finally 40 and 14 AGN in the \oi/\ha\ diagram.  
On the other hand, a large fraction of the  outflow candidates are grouped in the SF region. Only two galaxies lie above the K01 demarcation, one of them it is notably far away from this demarcation, in the AGN region (NGC\ 4388). As pointed in previous subsections, the classical interpretation of the diagnostic diagrams which attempt to separate between different sources of ionization is no longer valid without other extra parameters like the EW(\ha) or any other physical information about the source of ionization.
All candidates present EW(\ha) $>\, 3$ \AA\ in their nucleus. This means that according to the diagnostic diagrams and the EW(\ha) criterion, these galaxies are dominated by SF (15 of them) or AGN (1 weak and 1 strong).
 From the total candidates, 3/17 galaxies are catalogued as X-ray sources, 
NGC\,4676A \citep[$\log L_X = 39.2$,][]{Gonzalez-Martin2009}, NGC\,4388 \citep[$\log L_X = 42.45$,][]{Corral2014} and  NGC\,6286 \citep[$\log L_X = 40.6$][]{Brightman2011}. Altough only NGC\,4388 presents an X-ray luminosity greater than $\log L_X > 42$, the classical limit to be considered as an AGN. Indeed, this is the only target which outflowing material is compatible with being ionized by and AGN-driven ionization, based on the scheme described in Sec. \ref{select} (as indicated in Table \ref{table_candidates} and Figure \ref{galaxies_appendix}). In summary, our selection of highly inclined candidates to outflows seem to bias the sample towards outflows driven by star-formation in the vast majority. 

We should stress out that our selection bias the sample against early-type galaxies (as shown in the previous section), and this, by construction, excludes the detection of outflows in these galaxies that in their vast majority should be dominated by the presence of an AGN. In particular, we are excluding the detection of the recently classified as {\it Red Geysers}, a kind of object first reported by \citet{kehrig12}, and confirmed by \citet{cheung16}, and most probably associated with a weak AGN activity.


\begin{figure*}
\centering
\includegraphics[width=1\textwidth]{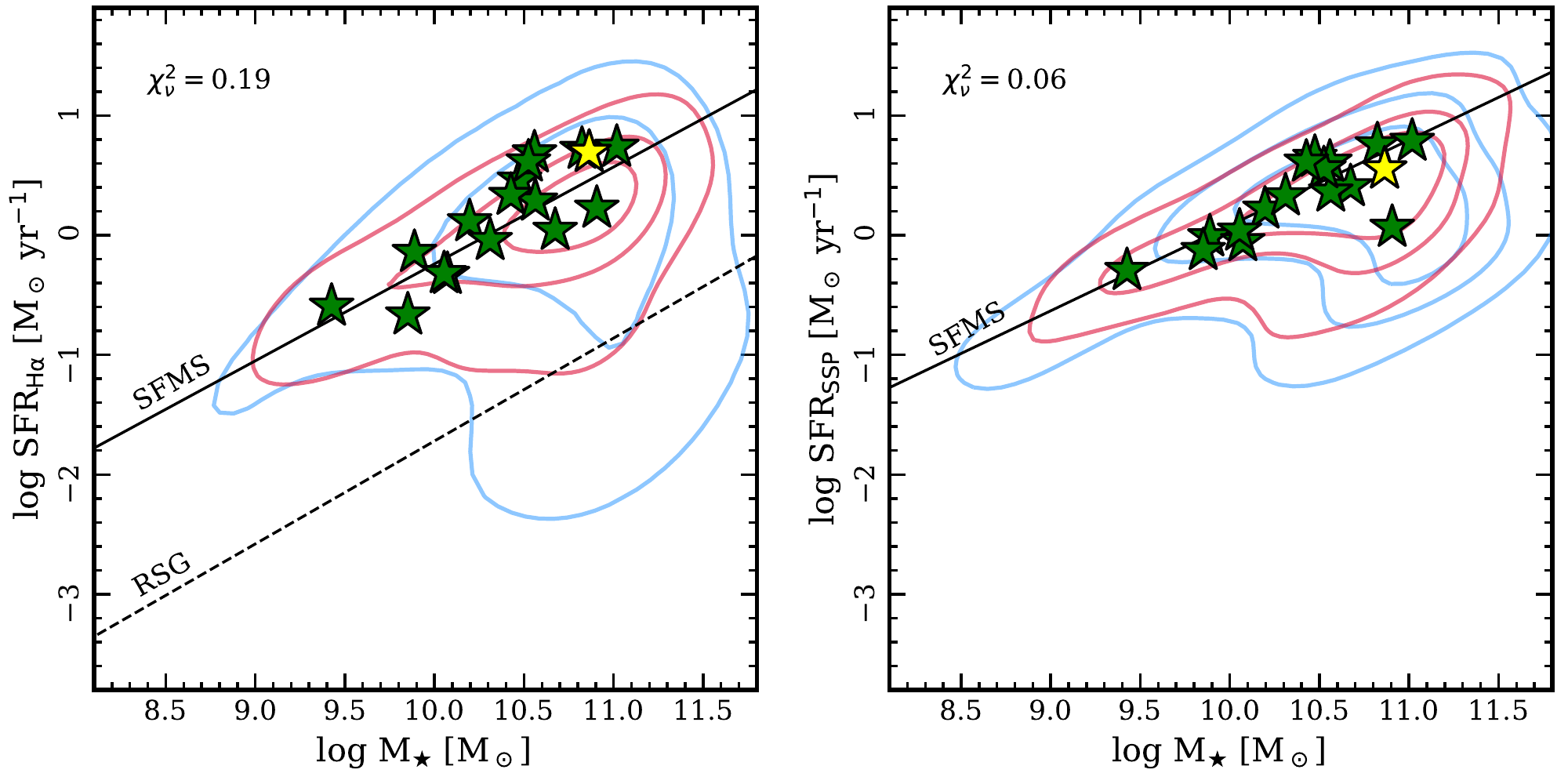}
\caption{Star formation rate derived from \ha\ ({\it left panel}) and estimated with the SSPs ({\it right panel}) vs. the integrated stellar mass  for the three sub-samples. Cyan and red contours 
enclose the 90, 68, and 34 \% of the total data in the CALIFA and the high inclination galaxies respectively. Green stars represent the outflow candidate galaxies. The yellow star represent the strong AGN, NGC\,4388, found in the candidates as shown in Fig.~\ref{BPT_sample}. 
The continuous and dashed black lines in the left panel correspond to the spatially resolved star formation main sequence (SFMS$_{\rm H\alpha}$) and the retired sequence of galaxies (RSG) derived by \citet{Cano2016}. The black line in the right panel correspond to the SFMS$_{\rm SSP}$ derived from the best fit for SF galaxies in the full CALIFA sample (EW(\ha) $>3$\AA\ and line ratios below the K01 curve) . The slope and zero point correspond to $0.71\pm0.02$ and $-7.05\pm0.20$ respectively. 
A chi squared test was applied for the candidates and the theoretical value given by the SMFS. The reduced chi squares is shown in the top left corner in each panel.}
\label{SFMS}
\end{figure*}


\subsubsection{Star formation rate vs. stellar mass}
\label{sec:SFMS}

\begin{figure*}
\centering
\includegraphics[width=\textwidth]{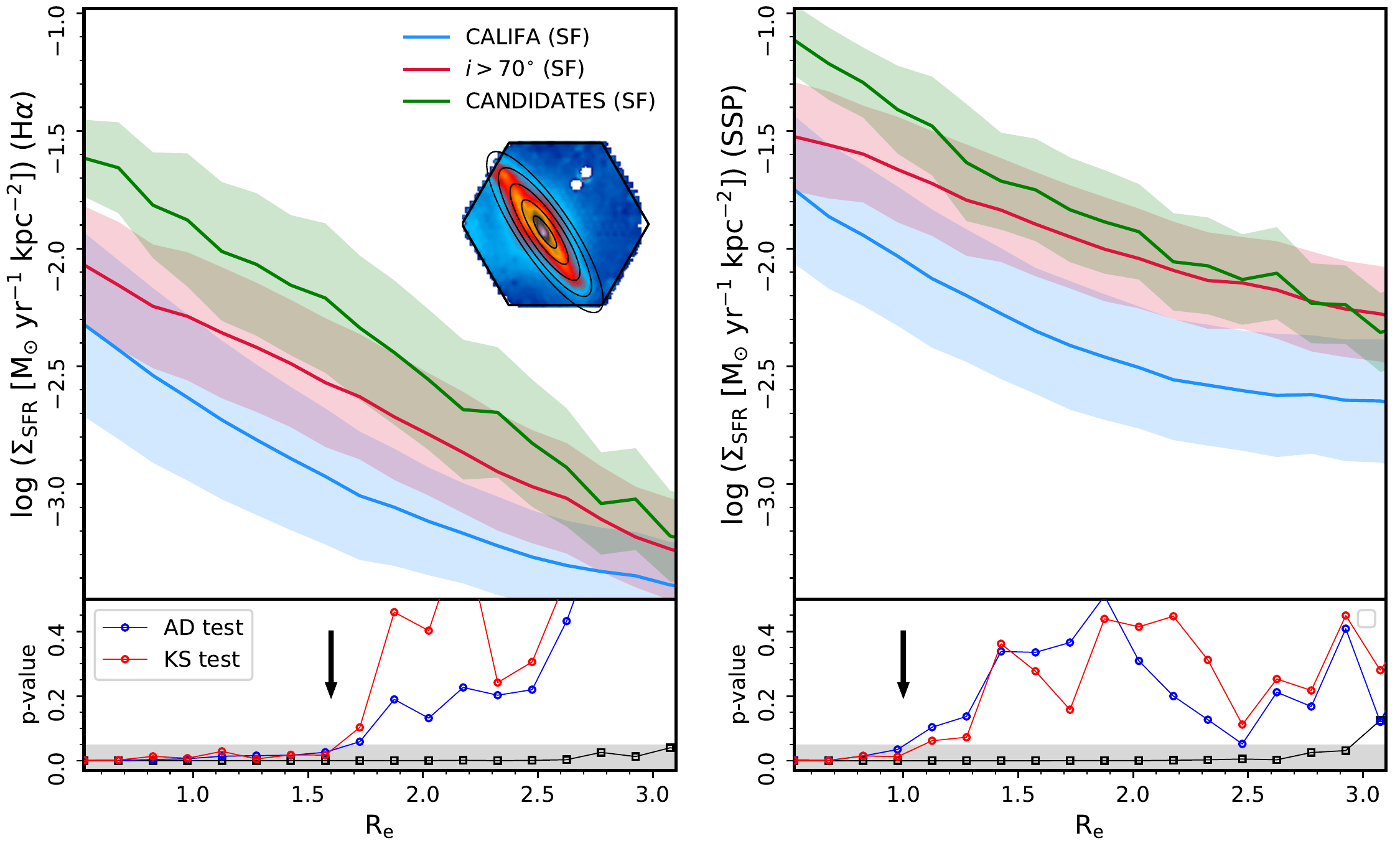}
\caption{Radial profiles of $\Sigma_{\rm SFR}$ for the three sub-samples estimated with \ha\ ({\it left panel}) and the SSP fitting analysis from {\sc Pipe3D} ({\it right panel}). In both panels  the green line represents the $\Sigma_{\rm SFR}$ profile for the outflow candidates (excluding the AGNs, see table \ref{table_candidates}), the red line represents the $\Sigma_{\rm SFR}$ radial profile for galaxies with high inclination catalogued as SF. The cyan line represent the radial profile for the low inclined galaxies (CALIFA subsample) catalogued as SF. We define SF as those galaxies with nuclear line-ratios lying below the K01 curve. 
Shadow regions represent the standard deviation at each bin of the effective radii. The bottom insets in both panel represent the A-D and K-S tests at each bin of the effective radius. The blue and red connected points represent the p-value obtained at each bin of the effective radius for the candidates and the high inclined galaxies for both statistical test. The black connected points represent the p-value obtained at each bin of the effective radius for the candidates and the CALIFA sub-sample. We have plotted the resulting p-values up to 0.5, larger p-values result obviously in the { acceptance} of the null hypothesis.
The black arrow in both insets represent the transition point where the null hypothesis  goes from being rejected to  being accepted. In the left inset this occur at ${\rm R}\,=\,1.6\,\rm{R_e}$ while in the right  plot occurs at ${\rm R}\,=\, 1\,\rm{R_e}$. An example of the ring segmentation performed  to calculate the radial distribution of $\Sigma_{\rm SFR}$ is shown  in the {\it left panel} as an illustration.
 } 
\label{Radial_profiles_ALL}
\end{figure*}

In the previous section we show that most of our candidates to host an outflow present ionized gas in their central regions dominated by star-formation (15 of 17). We explore in this section if this star-formation is more intense than the one of the other two sub-samples.

The SFR is a measurement of how much mass in stars is formed during a period of time. Star formation bursts create stars in a wide range of masses following a certain initial mass function \citep[e.g.,][]{Salpeter1955,Kroupa2001,Chabrier2003}, but only the massive ones will dominate the production of ionizing photons ($>$13.6 eV)  during a short period of time, $\sim$4 Myr. A common method to estimate the SFR in the optical range is through the luminosity of \ha\ \citep[$\mathrm{SFR}=7.93\,\times\,10^{-42}L_{\mathrm{H}\alpha}$;][]{Kennicutt1998}. This method requires that the measured \ha\ flux is produced only by SF process, which is not necessary true in the presence of an AGN, shocks or other ionization sources \citep[e.g.,][]{Catalan2015}. 
 To derive $L_{\mathrm{H\alpha}}$, we integrate the observed \ha\ flux, and after correction by dust attenuation using the extinction law by \citet{Cardelli1989}, assuming the case B of recombination \citep[e.g.,][]{Osterbrock} and using the cosmological distance for each galaxy. We applied the \citet{Kennicutt1998} law to transform $L_{\mathrm{H\alpha}}$ into SFR$_{\mathrm{H\alpha}}$. In this estimation we ignored the contribution of other sources of ionization. {However, as shown by \citet{Catalan2015} and \citet{sanchez17b} their effects are limited. Nevertheless, in retired galaxies dominated by old stellar population, this relation must be considered just as a linear transformation between the $L_{\mathrm{H\alpha}}$ to SFR.  

In Fig.~\ref{SFMS} we show the well known relation between the SFR and the integrated stellar mass \citep[e.g.,][]{Brinchmann2004, Salim2007,Noeske2007}. In this figure we plot both  the  SFR$_\mathrm{H\alpha}$ and SFR$_\mathrm{SSP}$ (explained in Sec.~\ref{ana_ssp}) for the  outflow candidates and the other two sub-samples.
In both panels the CALIFA sub-sample is distributed in a bimodal sequence shown by the blue contours, one comprising active star-forming galaxies, the so called star formation main sequence \citep[SFMS, e.g.,][]{Brinchmann2004, Salim2007}, and  the other the  passive or retired sequence of galaxies (RSG). These sequences have been previously studied spatially resolved for the CALIFA sample  \citep[e.g.,][]{Cano2016}. The high inclination galaxies are distributed around the SFMS with some galaxies falling in the RSG and the green valley. On the other hand, the outflow candidates are distributed around the SFMS regardless of the calibrator used to estimate the SFR, i.e., no excess is evident.
This result has been previously noticed by \cite{Ho2016} in a sample of outflows selected from the SAMI survey. Figure~\ref{SFMS} also  shows evidence that using the full optical extension of  galaxies, no excess in the SFR of the candidates is appreciated as it would be  expected if outflows are driven by strong periods of SF. 
The  outflow candidates seem to be part of the normal star-forming galaxies as revealed by the $\chi^2$ test.
Although it seems that outflows are preferentially located along the SFMS, their location in this diagram does not seem to define if a galaxy hosts or not an outflow.
Recent studies have pointed out that the local concentration of the SFR might play an important  role when driving outflows \citep[e.g.,][]{Ho2016}. In other words, the SFR surface density may be a better parameter instead of the integrated SFR to trace or regulate the presence of outflows. 

\subsection{Radial profiles of SFR surface density}
\label{sec:rad}

Early studies in local starburst galaxies and high--z Lyman break galaxies, have evidenced that outflows are ubiquitous in galaxies with SFR surface densities ($\Sigma_{\rm SFR}$) larger than  $10^{-1} \,\mathrm{M}_\odot \mathrm{yr}^{-1} \mathrm{kpc}^{-2}$ \cite[e.g.,][]{Heckman2001,Heckmanetal2002}. Based on these results this value has been adopted in the literature as a canonical threshold for outflows.

Motivated by these results, and the results from the previous section, we proceed to estimate the radial distribution of the $\Sigma_{\rm SFR}$. One of the great advantages of IFS is its capability to study the spatially resolved properties of galaxies, like $\Sigma_{\rm SFR}$, instead of deriving it averaged across the entire optical extension of galaxies like it was done in  previous analysis. 
For example, \citet{Kennicutt1998} used the area within the isophotal radius of the galaxies (D$_{25}$=2R$_{25}$) to estimate $\Sigma_{\rm SFR}$ ($={\rm SFR}/\pi {\rm R}_{25}^2$). Other authors have adopted the effective radii to estimate the area of the galaxies $\sim \pi \mathrm{R_{\rm e}^2}$ \citep[e.g.,][]{Lundgren2012,Ho2016} or it has been determined by imposing the Schmidt-Kennicutt law \citep[SK-law,][]{Kennicutt89}. In some cases in which it was possible to estimate the size of the starburst region (few hundreds of pc) it was adopted as the proper area where star formation is detected \citep[e.g.,][]{Wood2015}. { These differences in the procedure adopted to derive the  $\Sigma_{\rm SFR}$ introduce clear uncertainties in the absolute scale of the proposed canonical threshold described before.}

In our case we estimate the SFR derived from \ha\ and the SSPs at different galactocentric elliptical rings, following the position angle and ellipticity of the object. Then we divide each region by the physical area of the corresponding ring, corrected by the inclination angle, to finally obtain the radial distribution of $\Sigma_{\rm SFR}$ for each galaxy. We selected annular rings of 0.1 R$_{\rm e}$ width, up to 3 R$_{\rm e}$  c.f., Fig.~\ref{Radial_profiles_ALL}.
In addition, we estimate the $\Sigma_{\rm SFR}$ with the SFR derived from the SSP fitting analysis (SFR$_\mathrm{SSP}$), as described in Sec.~\ref{ana_ssp}. This method has the great advantage that it does not depend on the physical properties of the ionized gas. However, the SFR$_\mathrm{SSP}$ is only estimated where stellar continuum is detected. This means that  SFR$_\mathrm{SSP}$ traces pure SF with no contamination, at the penalty of a lower precision \citep[due to the limitations of the SSP-fitting procedure,][]{Pipe3D_I}. {\sc Pipe3D} estimates the SFR$_\mathrm{SSP}$ for the assembled mass in the last $\Delta t = 32 \mathrm{\,Myr}$, as described in Sec.~\ref{ana_ssp}. We adopted the same annular rings for this complementary estimation of the star-formation density.
\begin{figure*}
\centering
\includegraphics[width=1\textwidth]{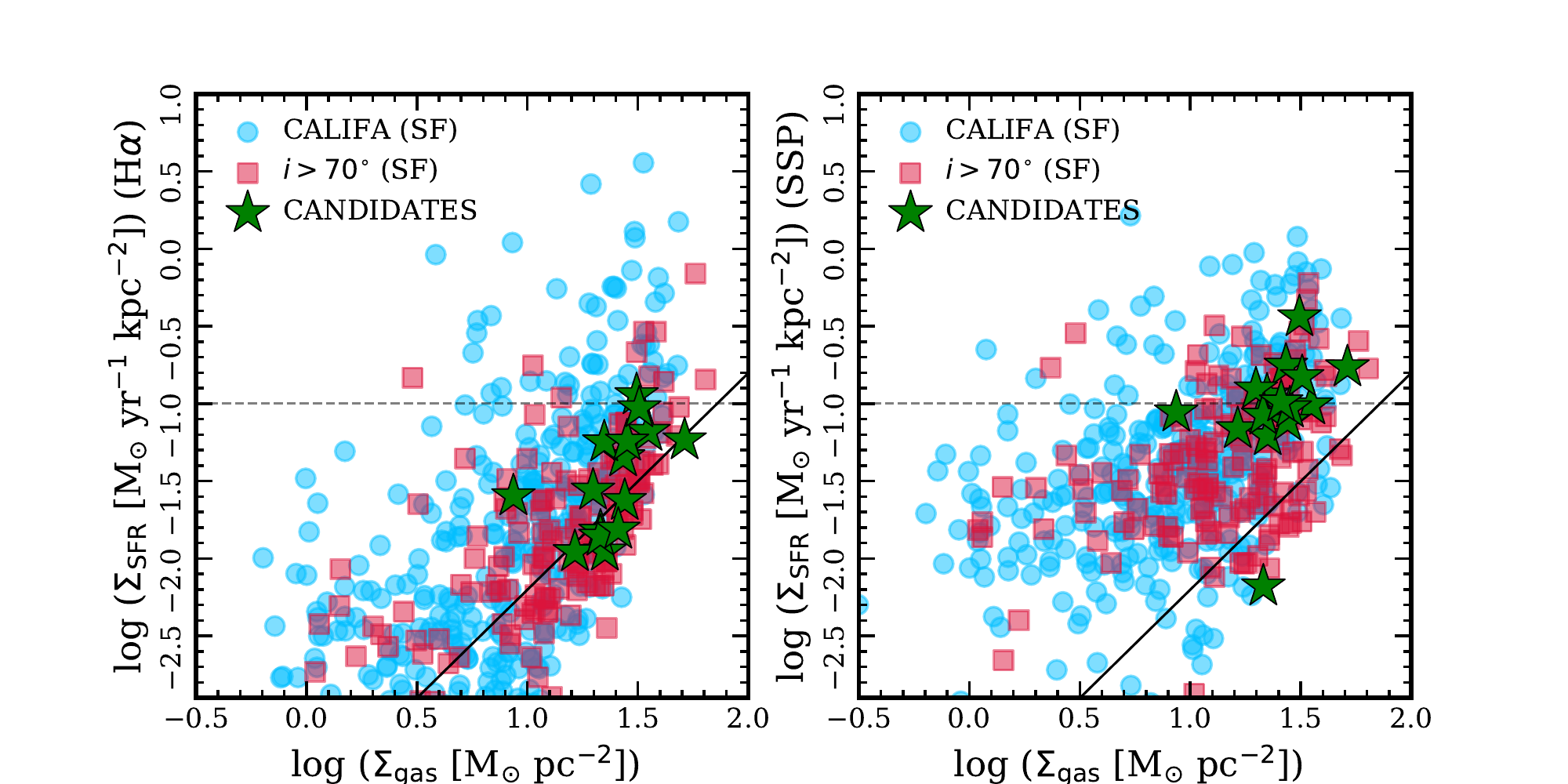}
\caption{$\Sigma_{\rm SFR}$ vs. $\Sigma_\mathrm{gas}$
for the individual galaxies in three sub-samples; {\it left panel} shows the $\Sigma_{\rm SFR}$ estimated with \ha; {\it right panel} shows the $\Sigma_{\rm SFR}$ estimated with the SSP fitting analysis. The cyan dots represent the SF galaxies in the CALIFA sub-sample. The red squares represent the SF galaxies in the high inclined sub-sample and the green stars the candidates (excluding the AGNs, see Table~\ref{table_candidates}). The green stars represent the outflow candidate galaxies. The horizontal dashed line represent the canonical threshold expected for outflows  ($\log \Sigma_{\rm SFR}$=1.0). The black straight line represent the \citet{Kennicutt1998} relation.
 } 
\label{SK}
\end{figure*}

Figure~\ref{Radial_profiles_ALL} shows the radial profiles of $\Sigma_{\rm SFR}$ estimated based on the  \ha\ flux and the SSP fitting analysis. These plots were constructed by taking the average value of the $\Sigma_{\rm SFR}$ for each galaxy in each radial bin, for each sub-sample of galaxies. The inner 0.2 ${\rm R_e}$ ($\sim 500$ pc) are unresolved due to the PSF size \citep[FWHM$\sim 2.4\arcsec$][]{CALIFA3}. Therefore, any trend below this inner region should be taken with care.
In this plot we considered only galaxies dominated by SF in each sub-sample (i.e., those galaxies with line ratios below the K01 curve in Fig. \ref{BPT_sample}). Adopting this criterion for the high and low inclined and outflow candidate galaxies, the sub-samples are limited to 412, 158 and 16 galaxies, respectively.

The radial profiles show that in both cases the candidate galaxies present, on average, higher values of the $\Sigma_{\rm SFR}$, at least in the inner regions, when compared to the other two sub-samples. The $\Sigma_{\rm SFR}$ estimated with the SSP method has in general larger values in comparison with the estimated with \ha\ because of the different time scales of both methods. The SFR based on \ha\ traces the SF in the last $\sim$4 Myr, while the SFR estimated with the SSPs traces the SF in the last 32 Myrs.
Regardless of the method used to estimate the SFR, it is clear that there is a trend in the outflow candidates to present larger values of the $\Sigma_{\rm SFR}$ in the innermost regions that in the outermost. 
This comparison between both estimators of the SFR exhibit that the observed excess is not due to a possible contamination by an extra source of ionization of \ha.

In order to quantify how significant is the difference in the radial distribution of the $\Sigma_{\rm SFR}$ of the candidates in comparison the two other samples, we performed an  Anderson-Darling test \citep[A-D,][]{Numerical_Recipes,Feigelson2012}. In contrast with the K-S test which tends to be more sensitive  to differences in the central regions of the distributions, the A-D test is more sensitive to differences also in the outermost regimes of the distributions.
We applied the A-D test at each radii, comparing the distributions of $\Sigma_{\rm SFR}$ for the candidates and the high inclination sub-samples, and the  outflow candidates and the low inclination galaxies. The null hypothesis in both cases is that at each radius both distributions of $\Sigma_{\rm SFR}$ are sub-samples from the same population. The results of this statistical tests are shown in the bottom insets of Fig.~\ref{Radial_profiles_ALL}. By adopting a significance level of 5\% we see that in the inner radii, the resulting p-value for the candidates and the high inclined galaxies is clearly below the significance level. This is still significant at a 1\% level. Although the radius at which the null hypothesis is rejected depends on the calibrator used to estimate the SFR, it is certainly clear that below $1 \,\mathrm{R_e}$ both distributions seems to be different. Indeed at this point the $\Sigma_{\rm SFR}$ profile presents a clear break and steepening most probably due to the outflow contribution inside this region.
As we go to  outer regions in the galaxies, the outflow candidates follow the same behaviour of the high-inclination galaxies. On the other hand, the p-values for the  outflow candidates and the low-inclination CALIFA sub-sample show that both distributions are different at any radius.

It is  interesting to note the low values of the  $\Sigma_{\rm SFR}$ that outflow candidates have, with values always lower $\log \Sigma_\mathrm{SFR}<-1.0$, which is below the canonical value expected for outflows, described before. We will explore this result in more detail in the next section.

\subsubsection{$\Sigma_{\rm SFR}$ vs. $\Sigma_\mathrm{gas}$: The K-S law.}
\label{sec:sig_SFR}
It has been frequently adopted the density of SFR as the main parameter that controls the production of an outflow \citep[e.g.,][]{Kennicutt1998,Heckmanetal2002}.
As larger the $\Sigma_{\rm SFR}$ is, more concentrated would be the energy released by the SN explosions and therefore the over pressured cavity would expand until large scale galactic winds  are driven giving rise to an outflow \citep[e.g.,][]{Heckman1990}.
To achieve high values in the $\Sigma_{\rm SFR}$ it is needed a high 
SFR concentrated in small regions (hundreds of pc). A large SFR  is reflected in a large fraction of gas that is transformed into newborn stars, from which only the massive ones (O-B stars) will contribute to the formation of the winds required to produce outflows. 

It is well known that there is a tight correlation between the $\Sigma_{\rm SFR}$ and the $\Sigma_\mathrm{gas}$ (molecular and atomic) content in galaxies \citep[the so-called K-S law,][]{Kennicutt89}. Although we are not able to measure directly the gas fraction, due to the lack of CO and HI observations for all galaxies in our sample, it is still possible to have an estimation of the molecular gas content via the dust extinction. 
Following \citet{SFSanchez2017}, we proceed to estimate the gas content via the  extinction $A_V$:
\begin{equation}
\Sigma_\mathrm{gas}=15\frac{A_V}{\rm{mag}} \:[\mathrm{M_{\odot}pc^{-2}}]
\label{eq1}
\end{equation}
This relation presents a scatter of $\sim 0.3$ dex when compared with CO measurements \citep[e.g.,][]{Galbany2017}, as explored in detail by Barrera-Ballesteros et al. (in prep.), based on the results from the EDGE-CALIFA survey \citep{bola13}. 

Figure~\ref{SK} shows 
the relation between the SFR surface density, estimated with both \ha\ and the SSP analysis, and the molecular mass density estimated from Eq.~\ref{eq1}, averaged within the central regions (${\rm R}\,<\,0.4{\rm R_e}$) of the 
individual galaxies of our three sub-samples.
If we focus in the left panel, we observe that the candidates are basically concentrated in the region of galaxies with SF around the K-S relation. On the other hand, the high inclination and the CALIFA sub-samples are distributed in a cloud, narrower for the first, also around the K-S relation. The scatter is larger than the usually reported for the K-S law, most probably due to the rough estimation of the gas density \citep[as already noticed by][]{sanchez17b}. 
In this panel we observe that only one of the candidate galaxies surpass the threshold of $10^{-1} \, \mathrm{M}_\odot \mathrm{yr}^{-1} \mathrm{kpc}^{-2}(=\Sigma_\mathrm{SFR,threshold})$. If we now focus on the right panel, we observe that a large fraction of the candidates are concentrated in a small region close the canonical value. Indeed $\sim$95\% of the candidates present SFR surface densities larger than 
 $10^{-1.5} \, \mathrm{M}_\odot \mathrm{yr}^{-1} \mathrm{kpc}^{-2}$. The galaxies depart from the canonical location of the K-S law in the right panel, mostly due to the different time scale sampled by the SFR derived  from the SSP analysis.
 As indicated before, the SSP analysis traces the SF in a longer period of time, (32 Myr and 4 Myr, respectively). Starbursts have typical time-scales of $<100$ Myr \citep[e.g.,][]{Leitherer2001}. This means that using \ha\ as calibrator to estimate the SFR we only measure the recent SF ($\sim 4$ Myr), while adopting the SFR of the SSP, we would be measuring an important fraction of the SF produced during the life time of a typical star-burst.
 
Finally, it seems that the  molecular mass density  must be also larger in galaxies hosting outflows, in addition to the $\Sigma_{\rm SFR}$. 

\section{Results and Discussion}
\label{sec:res}

We have explored the ionized gas properties for all galaxies from the full CALIFA sample to investigate the presence of outflows in the local universe. We imposed a set of criteria in the morphology, on the physical properties of the ionized gas and in the continuum to select a sample of candidate galaxies  with a host outflow. The adopted criteria are: (i) highly inclined galaxies, (ii) detection of extraplanar ionized gas, (iii) identification of an enhanced line ratios along the semi-minor axis, (iv) a biconical, bipolar or a symmetric morphology in the extraplanar gas, (v) EW(\ha)$>$ 3 \AA\ in the outflow regions.

Our main result is that only 17 galaxies seem to host outflows which correspond to 8\% from the highly inclined galaxies (273)  and 2\% from the extended CALIFA sample (835 objects). This last fraction is similar to what was found by \citet{Ho2016} in the SAMI galaxy survey.
We find that the galaxies hosting an outflow are located in the range of high mass $\log \mathrm{\,M}_{\star}>9.5$. Although in low mass galaxies outflows are less frequent, their local impact might be stronger than in galaxies with higher potentials. 
 
The amount of outflows detected in the full CALIFA sample may be a consequence of the short life-time of these processes. The dynamical time scale of outflows in starburst galaxies and AGN driven winds is in the range 
$\sim\,1-10$ Myr \citep[e.g.,][]{Veilleuxetal2005}.
CALIFA samples galaxies in the Local Universe, in a range of redshift between 0.005 and 0.03.  This range translates into a range of time of $\Delta\,t_{age} =0.34$ Gyr. This means that if the 17 outflows detected in this sample are representative of the full sample, then it is expected on average one outflow every 20 Myr. So, it is still possible that all galaxies in the sample have suffered an outflow process in the past, but that these were not observed due to their short life-time. The detection of outflows with much lower $\Sigma_{\rm SFR}$ than anticipated and their random location along the SFMS  might also reflect the stochasticity of these processes.

Fig. \ref{galaxies_appendix} shows that the vast majority of our outflow candidates present shock-excited emission lines in the extraplanar gas. This is quantified in  Table  \ref{table_candidates}, by the sharing between different ionization sources of the extraplanar ionized spaxels: most of them are dominated by shock-like SF-driven winds (16/17) and only one is consistent to be an AGN-driven wind: NGC\,4388. Indeed, this is the only target which central ionization is clearly compatible with the presence of an AGN and with a strong X-ray luminosity. From this analysis we conclude that most of our selected outflow candidates are consistent with being driven by star-formation. However, there is still the possibility that a galaxy hosts both SF and AGN activity causing a mixing in the ionization \citep[e.g.,][]{Davies2014}, which will produce a complex distribution of points along the diagram. In addition, the high inclination might produce a strong nuclear obscuration and blur the signal of a possible AGN. This would affect the observed optical emission lines, locating them in the SF region in the considered diagnostic diagrams. So far, we cannot reject non of both possibilities. Indeed, we find two targets, IC\ 2247 and NGC\ 4676A, with a fraction of $\sim$30\% of their extraplanar ionization being compatible with ionization by an AGN based on our criteria. The former one has a mixed/composite ionization in the central region (AGN/SF), while the later has clear X-ray emission although it is not considered to host an AGN \citep[][, and references therein]{Wild2014}. 

We found that the global SFRs of the outflow candidates puts all of them along the active star formation sequence, and that there is no significant excess in the SFR. This is contrary to expectation if this parameter was the major driver for the presence of outflows.
Nevertheless, when we explore the spatial concentration of the SFR, we observe that on average, the candidates do present an excess in their $\Sigma_{\rm SFR}$, when compared with galaxies with SF activity but without evidence of outflows.  This excess in $\Sigma_{\rm SFR}$ is statistically significant for the innermost regions and it holds up to $\sim$1 ${\rm R_e}$ ($\sim$ 4 kpc at the average redshift of CALIFA). This constrains a spatial region, which is larger in size compared to the typical acting region of a starburst $10^{2.5}-10^{3.5}$ pc \citep[e.g.,][]{Lehnert1996}, where the outflows have a significant signature in the properties of galaxies. For ${\rm R}\,>\,1 \, {\rm R_e}$ the outflow candidates behave as normal SF galaxies  and within this region the $\Sigma_{\rm SFR}$ distribution steepens. 

Although on average the candidates do not seem to surpass the proposed canonical threshold for the star formation surface density $\Sigma_{\rm SFR} > 10^{-1} \,\mathrm{M}_\odot \mathrm{yr}^{-1} \mathrm{kpc}^{-2}$, when we analyse  the individual values of this parameter we observe that all of them lie close to this canonical value. Depending on whether it is used the \ha\ or the SSPs calibration for the SFR, they can surpass the canonical value  only in a few cases. In starburst galaxies, the IR luminosity  ($L_{\mathrm{FIR}}$) is used as a tracer of the SFR. Indeed, the calibrator used to estimate the canonical threshold  in starburst galaxies adopts the $L_{\mathrm{FIR}}$ that traces the dust heating due to stars of 10--100 Myr  \citep[e.g.,][]{Kennicutt1998}.
Due to fact that the SFR$_\mathrm{SSP}$ comprises larger periods of SF, compared with \ha, the SFR$_\mathrm{SSP}$ might be used as a better estimator of the SFR in outflows. The sampled time by this calibrator approaches to the dust re-emission time scales.

Our results suggest that the threshold limit in $\Sigma_\mathrm{SFR}$ might be more flexible and include galaxies with lower values, in a regime where normal SF spiral galaxies dominates, rather than extreme starbursts. If we go back to the initial studies in outflows, we see that this threshold is achieved only for starburst and high-z Lyman break galaxies, and not for normal disk galaxies which can present values of the star-formation density as low as $10^{-3}-10^{0}\, \Sigma_\mathrm{SFR,threshold}$ \citep[e.g.,][]{Kennicutt1998,Heckman2001,Kennicutt2007}. The fact that outflows are ubiquitous in galaxies that exceed  the proposed limit does not exclude the possibility to find outflows in galaxies with $\Sigma_\mathrm{SFR}< \Sigma_\mathrm{SFR,threshold}$.
Indeed, more recent studies have also pointed that this threshold is quite high for the bulk population of outflows \citep[e.g.,][]{Ho2016}.

We have also shown that not only the $\Sigma_{\rm SFR}$ is a key parameter to generate outflows, but it must be accompanied with large densities of molecular gas (i.e.,  $\Sigma_\mathrm{gas}$). We propose a region for galaxies hosting outflows in the K-S diagram:  
$\Sigma_{\rm SFR}>10^{-2} \,\mathrm{M}_\odot \mathrm{yr}^{-1} \mathrm{kpc}^{-2}$  with $\Sigma_\mathrm{gas}>10^{1.2} \, \mathrm{M}_\odot  \mathrm{pc}^{-2}$, in a central region of $\sim$1 kpc.
In summary, it is not only the presence of strong star formation rate concentrated in a small area, but also the presence of material to be ejected what seems to be needed to generate an outflow. 

However, although this seems to be a necessary condition, only 17 galaxies from the highly inclined galaxies present an outflow. This suggests that these are not the only key parameters in driving outflows. Even more, the candidate galaxies are in the high mass range, contrary to what is expected to an outflow  can escape from their local potential. This last is a critical result in the context of the implication in the evolution of the low mass galaxies. The lack of high spatial resolution in our data  and our selection of high inclined galaxies might be biasing our sample to galaxies hosting large scale outflows. 
These are the are only detectable in the limit of the spatial resolution of CALIFA. The implementation of the technique introduced in this paper to detect outflows in high spatial resolution surveys is necessary to confirm our results in a more unbiased way.    
Our search for outflows has been performed without any bias toward the detection of starburst galaxies. 
In our ``blind" classification process we have been able to select galaxies that are previously known to host outflows, like the MICE or UGC\,10043 \citep[e.g.,][]{Wild2014,Coba2017}.
Another method for detecting outflows that we have not explored in this work is trough the interstellar absorption-line \nai\ $\lambda\lambda 5890,5896$ \citep[e.g.,][]{Heckman2000,Chen2010}. Nevertheless,  to our knowledge, we have not excluded any previously reported outflow from our explored sampled. 
\section{Conclusions}
\label{sec:con}

{The main conclusions of the exploration of the presence of outflows in the complete sample of galaxies observed by the CALIFA survey are the following ones:

\begin{itemize}

\item The fraction of galaxies with clear evidence of outflows range between 2\% to 8\%, depending if we consider the full sample of galaxies with any possible evidence or those ones that fulfil all our selection criteria.

\item The properties of galaxies hosting outflows are similar to that of the non hosting ones in terms of their distribution along the CMD, mass, morphology and integrated SFR, when the comparison is restricted to galaxies of the same inclination.

\item Galaxies hosting outflows are distributed in a high mass range of $9.5<\log \mathrm{\,M}_{\star}<11$. 

\item Most of our outflow candidates are compatible with being driven by star-formation, based both on the dominant ionization in the central regions and their location in the diagnostic diagrams in comparison with demarcation described by \citet{Sharp2010}. Only in one case we see clear evidence of AGN-driven outflows (NGC\ 4388).

\item The highly-inclined galaxies hosting an outflow present a significant excess in the star-formation rate surface density in the central regions (${\rm R}\,<\,1\,{\rm R_e}$), when compared with the non hosting outflow ones, indicating that at least in these galaxies, outflows are mostly driven by a central increase in the SFR.

\item The galaxies hosting outflows in the CALIFA sample only marginally exceed the canonical threshold on the  $\Sigma_{\rm SFR}$, maybe because they present regular star-formation which yields lower values in the 
star-formation surface density, and therefore produce weaker outflows compared to those of starburst galaxies.

\end{itemize}

Our results indicate that outflows are less restricted to extreme star-formation events, either central or integrated, being more frequent events than anticipated. Further studies are needed to explore the outflows in galaxies with lower inclinations, where data with better spatial and spectral resolutions could break the confusion between the different ionization components \citep[e.g.,][]{Coba2017}, and over much larger samples, like the ones provided by the MaNGA survey \citep[e.g.,][]{ManGA}, to provide with better statistics. Even more, we need to explore in more detail the physical properties of the outflows themselves, only outlined in the current study, and focus on the detectability of these events in retired/early-type galaxies, mostly excluded in this analysis due the  imposed inclination selection.

}

\section{Acknowledgements}

CLC and SFS are grateful for the support of a CONACYT (Mexico) grant CB-285080, and funding from the PAPIIT-DGAPA-IA101217(UNAM), PAPIIT: IN103318
 and CONACYT: 168251  projects. ICG, SFS and CLC acknowledge support from DGAPA-UNAM (Mexico) grant IN11341.
{ CLC acknowledges CONACYT (Mexico) Ph.\,D. scholarship.}  JBH acknowledges the support of an ARC Laureate Fellowship from the Australian Government. LG was supported in part by the US National Science Foundation under Grant AST-1311862.

This study  uses data provided by the Calar Alto Legacy Integral Field Area (CALIFA) survey (http://califa.caha.es/).

CALIFA is the first legacy survey performed at Calar Alto. The CALIFA collaboration would like to thank the IAA-CSIC and MPIA-MPG as
major partners of the observatory, and CAHA itself, for the unique
access to telescope time and support in manpower and infrastructures.
The CALIFA collaboration also thanks the CAHA staff for the dedication
to this project.

Based on observations collected at the Centro Astron\'omico Hispano
Alem\'an (CAHA) at Calar Alto, operated jointly by the
Max-Planck-Institut f\"ur Astronomie and the Instituto de Astrof\'\i sica de
Andaluc\'\i a  (CSIC).

\bibliographystyle{mnras}
\bibliography{ref,CALIFAI,joss}
\begin{landscape}
\begin{table}
\centering
\caption{Main properties of the galaxies candidates to host outflows.
$^1$ NASA/IPAC Extragalactic Database. $^2$ HyperLeda. $^3$  Estimated from an isophotal analysis on the SDSS r-band images as described in \citet{walcher14}. $^4$ Estimated from the SSP fitting analysis. $^5$ Estimated over an area of $3''\times3''$ around the nuclear region.$^6$ According to the  \citet{Kewley2001} demarcation.$^7$ Estimated at 0.2 Re.$^8$ The interaction is refereed as if there are closer companions at the same redshift in the SDSS images or if present evidence of interaction.
$^9$ Excitation mechanism for the observed outflow. The fractions represent the contribution in the  ionization of the extraplanar gas (spaxels lying beyond { 5\arcsec} from the disk) by an AGN, shocks or DIG. These fractions takes into account simultaneously all the points lying above the \citet{Kewley2001} curves in each diagnostic diagrams.
The shock and AGN fractions was estimated by the amount of spaxels  with EW(\ha) $>3$  \AA\ lying at the right- and left-side respectively of the shock/AGN excitation bisector lines from \citet{Sharp2010}, in the three diagnostic diagrams. The DIG fraction was estimated from the amount spaxels in the extraplanar region with EW(\ha) $<3$  \AA\ in the three diagnostic diagrams. }
\begin{tabular}{l c c c c c c c c c c c c c}
\hline
\hline
 Object  & z$^{(1)}$ & Hubble $^{(2)}$ & $i^{(3)}$ & PA$^{(3)}$ & R$_e^{(3)}$ & $\log$ M$_\star^{(4)}$ & $\log$ SFR$^{(5)}$ & Nuclear$^{(5,6)}$ & $\Sigma_\mathrm{H\alpha}^{(7)}$ & $\Sigma_\mathrm{SSP}^{(7)}$ & $\Sigma_\mathrm{gas}^{(7)}$&Interacting$^{(8)}$&AGN: Shock:DIG$^{(9)}$\\
 &  &  Type & [$^{\circ}$]& [$^{\circ}$]& [kpc]& [M$_\odot$] & [M$_{\odot}$\,yr$^{-1}$] &  ionization& [M$_{\odot}$,yr$^{-1}$\,pc$^{-2}$] & [M$_{\odot}$\,yr$^{-1}$\,pc$^{-2}$] & [M$_{\odot}$,pc$^{-2}$]& (Y/N)& \% \\
 \hline

IC\,2101 &   0.0149 & Scd  & 79.0 & 55.1 & 4.0 & 10.5 &0.5 & SF  & -7.17 & -7.03 & 1.55 & N&19:81:0\\ 
IC\,2247 &      0.0143 & Sab  & 72.1 & 50.9 & 4.6 & 10.7 &0.0 & AGN/SF & -7.48 & -7.45 & 1.52 &N&29:69:2\\ 
MCG\,-02-02-040 &     0.0119 & Scd  & 76.9 & -40.6 & 2.6 & 10.2 &0.1 & SF  & -7.35 & -6.80 & 1.42&N&10:90:0 \\ 
NGC\,4676A &  0.0222 & Sdm  & 77.0 & -89.4 & 8.3 & 10.9 &0.2 & SF  & -7.84 & -8.18 & 1.35 & Y&32:67:1\\ 
NGC\,0216 &       0.0052 & Sd   & 71.1 & -59.9 & 1.3 & 9.4 &-0.6 & SF  & -7.60 & -7.05 & 0.94&N&4:96:0 \\ 
NGC\,6168 &      0.0088 & Sc   & 79.6 & 18.7 & 2.6 & 9.9 &-0.1 & SF  & -7.55 & -6.97 & 1.30&N&1:99:0 \\ 
UGC\,09113 &      0.0107 & Sb   & 72.4 & -32.8 & 2.4 & 10.1 &-0.3 & SF  & -7.96 & -7.21 & 1.37&N&1:99:0 \\ 
UGC\,09165 &      0.0177 & Sa   & 78.3 & -35.2 & 3.1 & 10.8 &0.7 & SF  & -7.10 & -6.77 & 1.70 &N&1:99:0\\ 
UGC\,10123 &      0.0126 & Sab  & 77.7 & -33.5 & 2.7 & 10.6 &0.3 & SF  & -7.63 & -4.76 & 1.44 &N&17:83:0\\ 
UGC\,10384 &      0.0167 & Sb   & 79.7 & 0.5 & 3.0 & 10.6 &0.7 & SF  & -4.66 & -4.38 & 1.01&N&10:90:0 \\ 
IC\,0480 &      0.0154 & Sc   & 82.3 & 76.7 & 3.5 & 10.3 &-0.0 & SF  & -7.97 & -7.15 & 1.21&N&0:100:0 \\ 
NGC\,5434B &      0.0190 & Sc   & 82.2 & -19.7 & 5.2 & 10.5 &0.6 & SF  & -7.29 & -7.17 & 1.45&Y&15:85:0 \\ 
UGC\,03539 &     0.0111 & Sc   & 83.4 & 25.6 & 2.8 & 10.1 &-0.3 & SF  & -7.89 & -7.15 & 1.33 &N&14:84:0\\ 
UGC\,10043 &      0.0074 & Sab  & 82.8 & 60.6 & 3.5 & 9.9 &-0.7 & SF  & -7.79 & -7.04 & 1.40&N&1:94:5 \\ 
NGC\,4388 &      0.0084 & SBb   & 66.3 & -2.4 & 3.5 & 10.9 &0.7 & AGN & -6.61 & -6.62 & 1.34& N&60:40:0\\ 
NGC\,6286 &      0.0183 & Sb   & 75.6 & -55.7 & 6.1 & 11.0 &0.7 & SF  & -7.02 & -6.87 & 1.51&Y&4:96:0 \\ 
MCG\,+11-08-25&   0.0136& Sab&54.6&23.7&5.0&10.4&0.3&SF& -7.09&-6.72&1.44&N&0:100:0 \\
\hline
\hline
\end{tabular}
\label{table_candidates}
\end{table}
\end{landscape}


\clearpage
\appendix
\section{Candidates galaxies}
\label{appendix:outlows}
In Fig.~\ref{galaxies_appendix} it is shown the spatially resolved line ratio maps and diagnostic diagrams for all the candidate galaxies with a host outflow listed in Table~\ref{table_candidates} comprising the same information shown for NGC\,6282 in Figs.~\ref{example_ic2101} and \ref{DD_ic0480}.

%
\begin{figure*}
\centering
\includegraphics[height=3.5in,width=\textwidth]{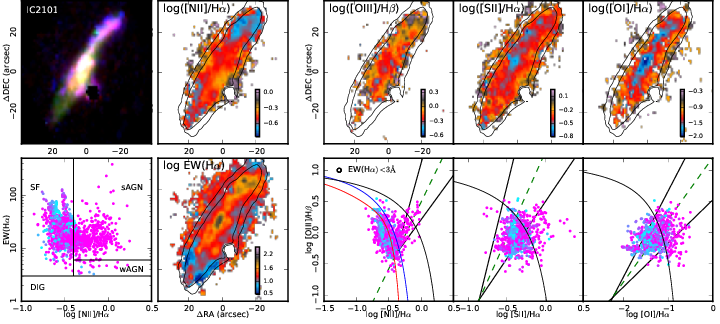}
\includegraphics[height=3.5in,width=\textwidth]{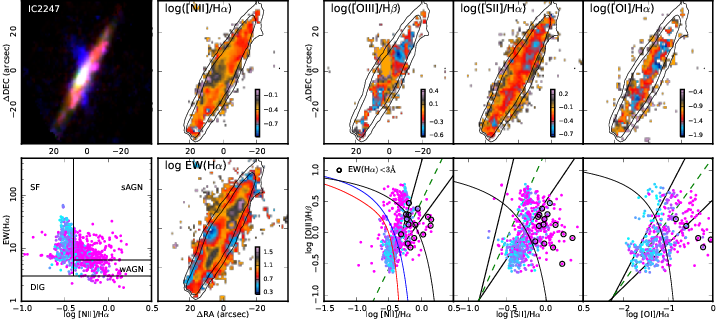}
\caption{Spatially resolved line ratios and diagnostic diagrams together the WHAN diagram for the candidates galaxies to host outflows listed in Table \ref{table_candidates}. In each panel it has included a false colour image of the galaxy ( green: V-band, red: \nii\ and blue: \oiii). The two black contours indicate the continuum level at 0.1 and 0.05 $\times 10^{-16}$ erg s$^{-1}$.
The meaning of the demarcation curves and the symbols are the same from Figs. \ref{example_ic2101} and \ref{DD_ic0480}. }
\label{galaxies_appendix}
\end{figure*}
\addtocounter{figure}{-1}
\begin{figure*}
\centering
\includegraphics[height=3.5in,width=\textwidth]{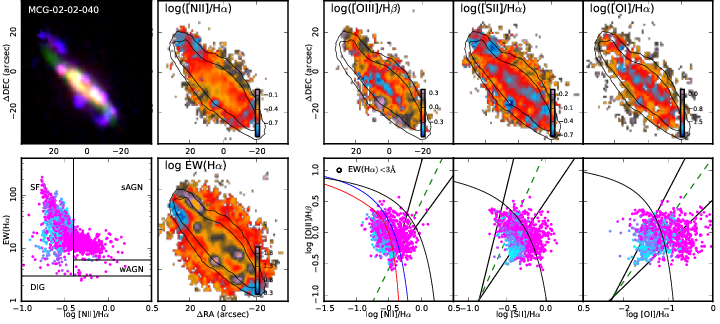}
\includegraphics[height=3.5in,width=\textwidth]{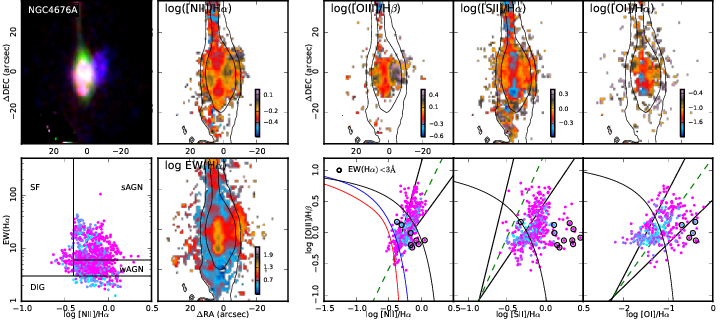}
\caption{({\it continue})}
\end{figure*}
\addtocounter{figure}{-1}
\begin{figure*}
\centering
\includegraphics[height=3.5in,width=\textwidth]{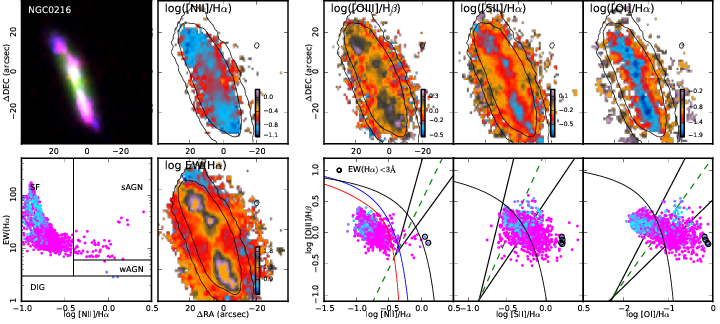}
\includegraphics[height=3.5in,width=\textwidth]{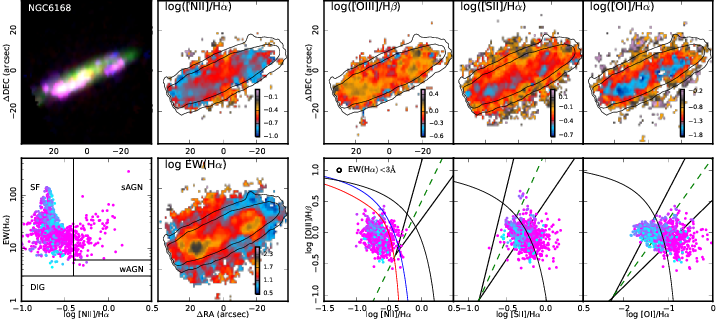}
\caption{({\it continue})}
\end{figure*}
\addtocounter{figure}{-1}
\begin{figure*}
\centering
\includegraphics[height=3.5in,width=\textwidth]{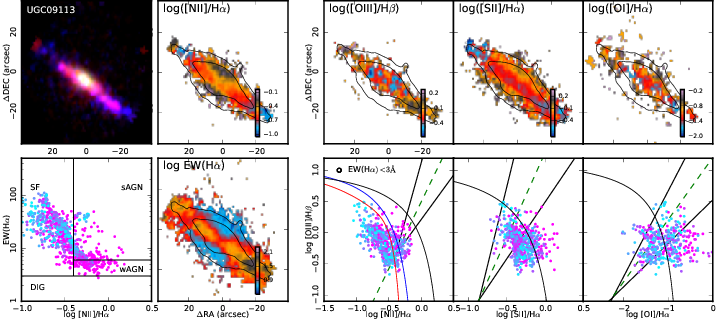}
\includegraphics[height=3.5in,width=\textwidth]{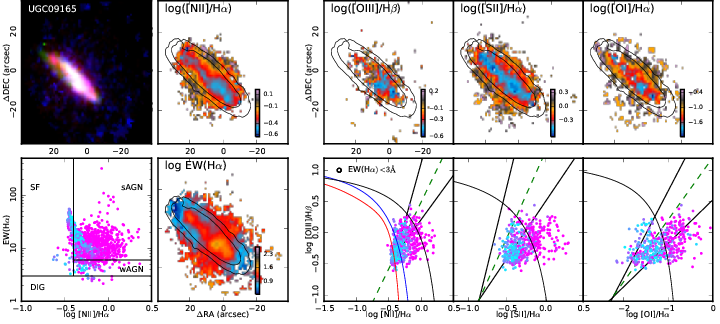}
\caption{({\it continue})}
\end{figure*}
\addtocounter{figure}{-1}
\begin{figure*}
\centering
\includegraphics[height=3.5in,width=\textwidth]{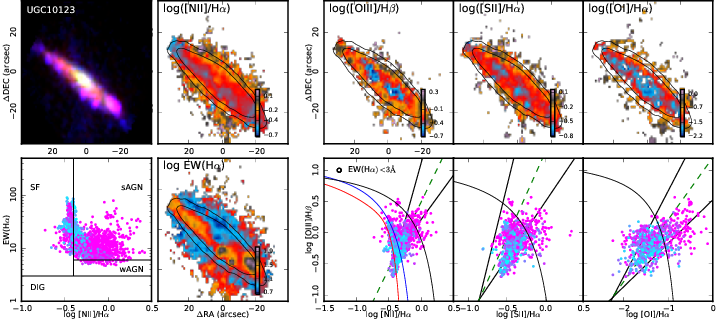}
\includegraphics[height=3.5in,width=\textwidth]{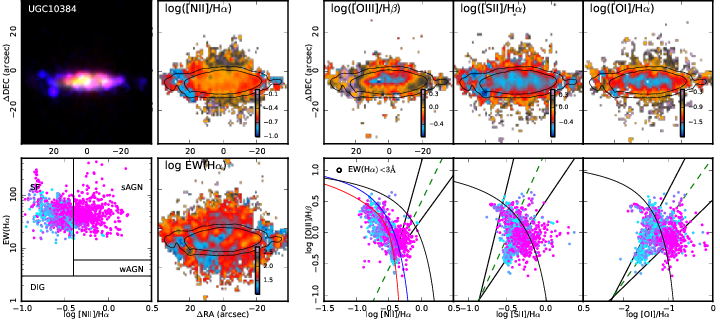}
\caption{({\it continue})}
\end{figure*}
\addtocounter{figure}{-1}
\begin{figure*}
\centering
\includegraphics[height=3.5in,width=\textwidth]{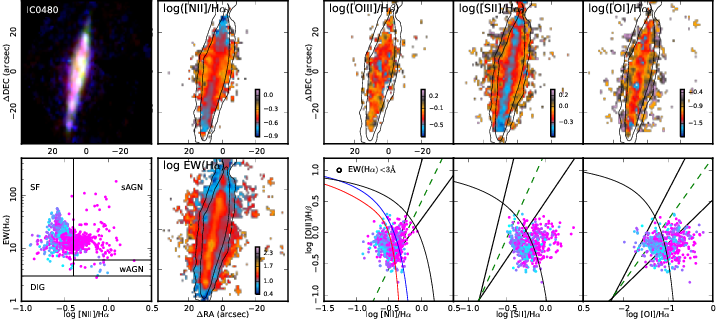}
\includegraphics[height=3.5in,width=\textwidth]{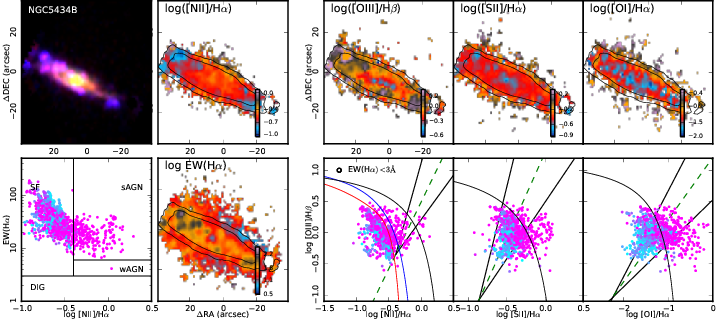}
\caption{({\it continue})}
\end{figure*}
\addtocounter{figure}{-1}
\begin{figure*}
\centering
\includegraphics[height=3.5in,width=\textwidth]{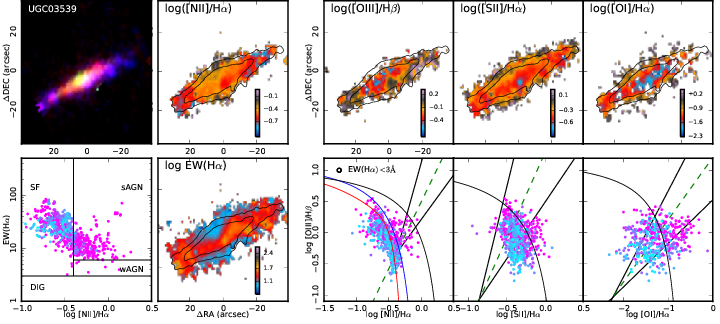}
\includegraphics[height=3.5in,width=\textwidth]{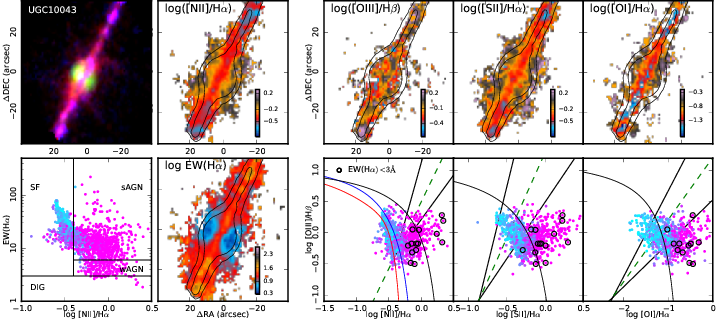}
\caption{({\it continue})}
\end{figure*}
\addtocounter{figure}{-1}
\begin{figure*}
\centering
\includegraphics[height=3.5in,width=\textwidth]{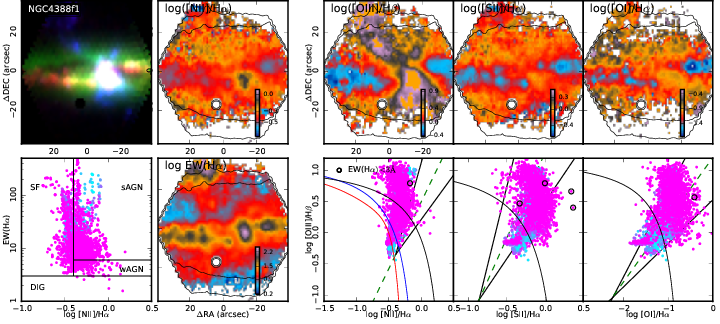}
\includegraphics[height=3.5in,width=\textwidth]{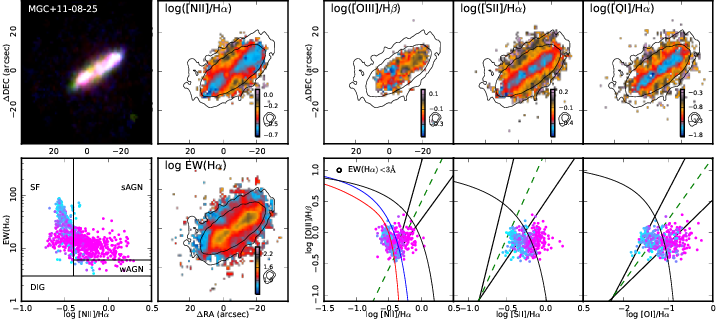}
\caption{({\it continue})}
\end{figure*}

\section{Galaxies not catalogued as outflows}
\label{appendix:no_outlows}

In Table~\ref{Tab:Not_outflow} we present the remaining galaxies with detected extraplanar ionized gas or some increase in the line ratios, but that were not classified as outflow candidates. Some of these galaxies present extraplanar  diffuse ionized gas (eDIG). In some cases the eDIG is dominated by HOLMES or post-AGB. In other cases the extraplanar gas presents EW(\ha) $>3$  \AA, but it is distributed in a continuous layer of ionized gas above the galactic disk. 
These ones might be ionized by leaking photons of \hii\ regions that scape from the disk. In galaxies with high star-formation rates, or starburst galaxies, a fraction of the ionizing photons can scape from the \hii\ regions without been absorbed. It has been suggested that this leaky \hii\ regions photons may scape to the diffuse ISM and the inter galactic medium and ionize regions of kilo parsec scales from the disk \citep[e.g.,][]{Ferguson1996,Hoopes2003,Wood2010,Martin2015}. 

It is important to emphasize that this work is not focused in the exploration of eDIG in general. There are other studies in the CALIFA survey that focuses in the analysis of DIG in all type of galaxies regardles of their inclination \citep[e.g.,][]{sign13,Lacerda2018}. Nevertheless we list in here those galaxies that might be probably confused with outflows, which do not imply these galaxies are the only ones with eDIG in the  CALIFA sample. It may be also possible that some of these galaxies could be re-classified as outflow candidates with better spatial and spectral resolution data.

\begin{table}
\caption{Galaxies with extraplanar  ionized gas, but not classified as outflows because they do not fulfil all the required criteria indicated in Sec \ref{select}. The fraction of spaxels in the extraplanar region compatible with being ionized by an AGN, SF-driven shocks, and old-stars, based on the scheme described in Sec. \ref{select} and Table~\ref{table_candidates} is included for reference.}
\begin{tabular}{l c}
\hline
\hline
Object&AGN:Shock:DIG\\
\hline
NGC\,0693&9:91:0\\
PGC\,0063016&1:99:0\\
UGC\,04730&23:77:0\\
UGC\,5392&6:94:0\\
NGC\,1677&1:99:0\\
NGC\,4149&2:80:18 \\
NGC\,5908&7:63:30 \\
MCG\,-01-01-012&0:50:50\\
NGC\,2480&8:92:0\\
NGC\,5402&0:100:0\\
NGC5439&0:100:0 \\
NGC\,6361&5:88:7 \\
UGC\,04550&12:88:0\\
UGC\,09262&2:98:0\\
UGC\,09665&0:100:0\\
IC\,2098&0:98:2\\
IC\,4582&3:97:0\\
NGC\,0681&0:72:28 \\
NGC\,1056&12:87:1\\
NGC\,5145&3:94:2 \\
IC\,1481&56:43:1\\
IC\,0540&30:14.:57\\
\hline
\hline
\end{tabular}
\label{Tab:Not_outflow}
\end{table}

\section{supplemental material}
Fig. \ref{fig:excluded} shows the same plots from Fig. \ref{galaxies_appendix} but for a galaxy not classified as an outflow candidate from Table \ref{Tab:Not_outflow}.
The remaining plots for the galaxies listed in Table \ref{Tab:Not_outflow} are available in the supplementary material for this article.
\begin{figure*}
\centering
\includegraphics[height=3.5in,width=\textwidth]{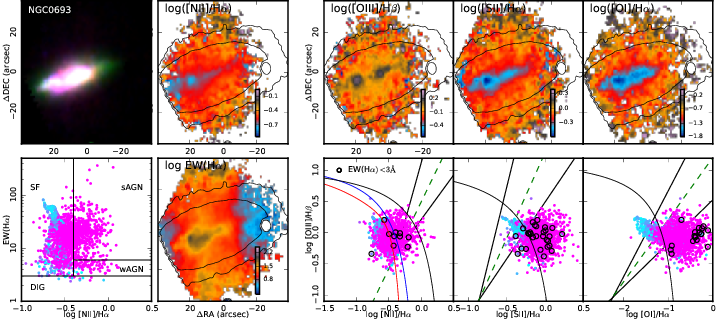}
\caption{Galaxy with detected extraplanar emission but not selected as outflow candidate. In each panel it has included a false color image of the galaxy (green: \ha\, red: \nii\ and blue: \oiii). The two black contours indicate the continuum level at 0.1 and 0.05 $\times 10^{-16}$ erg s$^{-1}$. { Black circles in the diagnostic diagrams indicate  EW(\ha) $<$ 3 \AA.}
The meaning of the demarcation curves and the color code of the symbols are the same from  Figs. \ref{example_ic2101} and \ref{DD_ic0480}. }
\label{fig:excluded}
\end{figure*}

\bsp	
\label{lastpage}
\end{document}